\documentclass[paper]{JHEP3} 

\usepackage{epsfig,multicol,multirow}
\usepackage{amsmath,subfigure,latexsym,amssymb,cite}
\usepackage{here}

\newcommand\fverb{\setbox\fverbbox=\hbox\bgroup\verb}
\newcommand\fverbdo{\egroup\medskip\noindent%
			\fbox{\unhbox\fverbbox}\ }
\newcommand\fverbit{\egroup\item[\fbox{\unhbox\fverbbox}]}
\newbox\fverbbox

\newcommand {\beq} {\begin{equation}}
\newcommand {\eeq} {\end{equation}}
\newcommand {\beqa}{\begin{eqnarray}}
\newcommand {\eeqa}{\end{eqnarray}}

\newcommand {\tr}{{\rm tr\,}}

\newcommand{\1}{\mbox{1}\hspace{-0.25em}\mbox{l}}

\def\theenumi{\roman{enumi}}

\allowdisplaybreaks

\title{Realizing three generations of the Standard Model fermions
in the type IIB matrix model}

\author{Hajime Aoki,$^{a}$
Jun Nishimura${}^{bc}$ and Asato Tsuchiya${}^{d}$
\vspace*{0.5cm} \\
\llap{$^a$}Department of Physics, Saga University, 
Saga 840-8502, Japan \\
\llap{$^b$}Department of Particle and Nuclear Physics,\\
Graduate University for Advanced Studies (SOKENDAI),\\
Tsukuba, Ibaraki 305-0801, Japan\\
\llap{$^c$}KEK Theory Center, 
High Energy Accelerator Research Organization,\\
Tsukuba, Ibaraki 305-0801, Japan\\
\llap{$^d$}Department of Physics, Shizuoka University,\\
836 Ohya, Suruga-ku, Shizuoka 422-8529, Japan
\vspace*{0.5cm} \\
\email{haoki@cc.saga-u.ac.jp, jnishi@post.kek.jp, satsuch@ipc.shizuoka.ac.jp}}

\preprint{SAGA-HE-280\\ KEK-TH-1701}

\abstract{
We discuss how the Standard Model particles
appear from the type IIB matrix model,
which is considered to be
a nonperturbative formulation of 
superstring theory.
In particular, we are concerned with
a constructive definition of the theory,
in which we start with finite-$N$ matrices and 
take the large-$N$ limit afterwards.
In that case, it was pointed out recently that
realizing chiral fermions in the model is
more difficult than
it had been thought from formal arguments at $N=\infty$
and that introduction of a matrix version of the 
warp factor is necessary.
Based on this new insight, we 
show that two generations of the Standard Model fermions
can be realized by considering 
a rather generic configuration of
fuzzy ${\rm S}^2$ and 
fuzzy ${\rm S}^2 \times {\rm S}^2$ 
in the extra dimensions.
We also show that three generations can 
be obtained by squashing one of the 
${\rm S}^2$'s that appear in the configuration. 
Chiral fermions
appear at the intersections of the fuzzy manifolds
with nontrivial Yukawa couplings to the Higgs field,
which can be
calculated from the overlap of their wave functions.
}


\begin{document}

\section{Introduction}
\label{sec:introduction}

The Standard Model of particle physics is a very successful
theory in that it can describe all the phenomena up to 
the energy scale reachable so far by accelerators.
Yet it cannot be considered as a fundamental theory
since it does not include quantum gravity.
Also the Standard Model has quite an involved structure
with many parameters, which is expected to be explained
by a fundamental theory like superstring theory.
Indeed there has been a lot of work in this direction with 
remarkable
success.
Superstring theory includes quantum gravity consistently,
and it is a simple theory with only one scale parameter.
Despite the simpleness of the theory,
one can find perturbative vacua,
which give rise to the Standard Model with some extra exotic particles.
However, there are some serious problems as well.
It seems highly nontrivial to fix all the moduli of the
perturbative
vacua although there are some new ideas such as the flux compactification.
Moreover, it is known that
there actually exist 
tremendously many perturbative vacua,
which is a situation commonly referred to 
as the string landscape nowadays.
From this point of view, one cannot even explain why we live in
a four-dimensional space-time since perturbative vacua
can have various space-time dimensionality less than or equal to ten.

Of course, all these problems might be simply 
because superstring theory has been studied
essentially in perturbation theory
including, at most,
some nonperturbative effects represented by the existence
of D-branes.
Therefore a different picture might emerge
if one studies the theory in a completely nonperturbative framework.
As a well-known example, nonperturbative studies of QCD by
lattice gauge theory explained important
low-energy dynamics such as confinement of quarks, which can never
be understood in perturbation theory.
The hadron mass spectrum has been reproduced
accurately by Monte Carlo calculations based on the
lattice gauge theory,
and such a method has been playing a crucial role in studying various
properties of hadrons.
Likewise it is possible that the compactification of extra six 
dimensions can be understood as a nonperturbative effect in
superstring theory, and that the involved structure of the Standard
Model and its parameters can be understood from a rather simple
structure in the extra dimensions.

%
%
As a nonperturbative formulation of superstring theory,
we consider the type IIB matrix model \cite{IKKT}, which consists
of 10 bosonic $N\times N$ Hermitian matrices $A_\mu$ ($\mu =  0 ,
\ldots , 9$) and
16 fermionic $N\times N$ Hermitian matrices $\Psi_\alpha$
($\alpha = 1, \ldots , 16$).
The action of the model can be formally obtained 
from that of ten-dimensional ${\cal N}=1$ SU($N$) 
super Yang-Mills theory by dimensional reduction.
The Yang-Mills coupling, which is the only parameter of the model,
becomes just a scale parameter after the dimensional reduction
since it can be absorbed by appropriate rescaling of the matrices.
The type IIB matrix model has a direct connection to perturbative type IIB
superstring theory, 
but it is expected to describe
the unique nonperturbative theory of superstrings 
underlying the duality web of various perturbative formulations.
The matrix size $N$ 
corresponds to the
number of sites
in the lattice gauge theory, which makes the dynamical degrees of
freedom in the system finite.
By taking the large-$N$ limit in an appropriate manner, 
one obtains
a constructive definition of superstring theory.
Nowhere in the definition does one have to make a perturbative
expansion,
hence it is a nonperturbative formulation.
A particularly interesting aspect of the type IIB matrix model
is that 
space-time is treated as a part of dynamical degrees of freedom 
in the bosonic matrices $A_\mu$. 
It is therefore possible that four-dimensional space-time appears
dynamically.

For more than fifteen years since its proposal,
the type IIB matrix model has been studied in its Euclidean version,
which can be obtained by making a ``Wick rotation'' $A_0 = i A_{10}$.
This is fine when one calculates the interactions between D-branes
at the one-loop level, for instance, but the physical meaning of
the ``Wick rotation'' 
is not clear
at a fully nonperturbative level unlike in quantum field theory.
The Euclidean version has been studied intensively, nevertheless,
because it has a finite partition 
function \cite{Krauth:1998xh,Austing:2001pk}.
See refs.\cite{Nishimura:2011xy,Anagnostopoulos:2013xga}
for recent work, which suggests that 
spontaneous breaking of the SO(10) symmetry occurs in the Euclidean 
matrix model.

The Lorentzian version
remained untouched until recently since it
looked simply ill-defined
due to its non-positive-definite action.
However, ref.~\cite{Kim:2011cr} showed
that the Lorentzian version can be made well-defined nonperturbatively
by first introducing infrared cutoffs in both temporal and spatial 
directions, and then removing them in the large-$N$ limit
in such a way that the continuum and infinite-volume limits are taken.
The resulting theory has no parameters other than one scale parameter.
Moreover, it turned out that
one can extract a real-time evolution by taking an ensemble average
over matrix configurations,
which showed that (3+1)-dimensional expanding universe emerges
dynamically.
This provides a strong evidence that the Lorentzian version 
of the type IIB matrix model indeed describes the unique nonperturbative
theory of superstrings, and that the theory provides a natural explanation
for the origin of our 4-dimensional space-time.

The aim of the present work is to discuss whether the same theory
can also provide a natural explanation for the origin of the Standard Model.
In particular, the Standard Model has the following peculiar features:
\begin{enumerate}
\item The gauge interaction is governed by
${\rm SU}(3)\times {\rm SU}(2)\times {\rm U}(1)$, which is
a semi-simple group instead of being a simple one.
\item The matter contents are fermions, which couple to the
gauge fields in a characteristic manner.
Quarks couple to the SU(3) gauge field, while leptons do not.
Left-handed fermions couple to the SU(2) gauge field,
while right-handed ones do not.
The assignment of the U(1) hypercharge is quite involved.
\item The matter contents have three generations, which couple
to the gauge fields in exactly the same manner.
\item The fermions have Yukawa couplings to the Higgs field.
Due to the existence of three generations, 
the Yukawa couplings involve a lot of parameters, which can only 
be determined experimentally within the Standard Model.
\end{enumerate}
We discuss how these features can be realized in
the type IIB matrix model.

One of the biggest obstacles in obtaining the Standard Model
from higher-dimensional theories like superstring theory
is that the fermions should be chiral.
If one applies a naive dimensional reduction to 
higher dimensional theories,
fermions become vector-like in four dimensions.
In order to realize chiral fermions,
one needs to consider, for instance, 
(a) orbifolding, which amounts to imposing
a nontrivial identification in the extra dimensions,
(b) introducing intersecting/magnetized D-branes,
(c) introducing a nonzero Euler number in the Calabi-Yau compactification.

Realization of chiral fermions and the Standard Model
has been discussed also in the type IIB matrix model 
by various authors.
In ref.~\cite{Aoki:2002jt} an orbifolding condition was
imposed on the matrix configuration, and it was shown to 
give rise to a four-dimensional 
theory including chiral fermions.
See also 
refs.~\cite{Chatzistavrakidis:2010xi,Chatzistavrakidis:2012ah}
for related works.
Ref.~\cite{Aoki:2010gv} studies a matrix model in which
the Hermitian matrices $A_a$ ($a=4 , \ldots , 9$) 
in the extra dimensions are replaced by 
unitary matrices $U_a$ with an action obtained as a one-site
model of six-dimensional SU($N$) lattice gauge theory.
These matrices $U_a$ can have solutions representing
a 6d non-commutative torus
carrying magnetic fluxes.
Chiral fermions can be obtained in this background
if one uses a Ginsparg-Wilson Dirac operator, 
which has an exact modified chiral symmetry.
One can realize three generations of the Standard Model particles
by choosing the fluxes in the 6d torus appropriately.\footnote{As a 
closely related work, refs.~\cite{Abe:2012fj,Abe:2013bba}
discuss
realization of the Standard Model by toroidal compactification of
ten-dimensional ${\cal N}=1$ SU($N$) super Yang-Mills theory
with orbifold conditions.
Realistic CKM and PMNS matrices were obtained
by choosing the vacuum expectation values of 
such quantities as Wilson lines, the K\"ahler moduli, 
the complex-structure moduli and the dilaton \cite{Abe:2012fj}.}
The probability distribution for the appearance 
of the Standard Model and other phenomenological models
has been calculated in this setup \cite{Aoki:2012ei,Aoki:2013pba}.

While the above proposals attempt to realize chiral fermions
by modifying the model,
ref.~\cite{Chatzistavrakidis:2011gs} proposed 
to realize chiral fermions
in the original type IIB matrix model
based on the idea of intersecting 
branes \cite{Berkooz:1996km}, which
has been explored extensively in 
the phenomenological context \cite{Antoniadis:2000ena,%
Aldazabal:2000sa,hep-th/0007024,Ibanez:2001nd,%
Blumenhagen:2001te,hep-th/0107143,%
Cvetic:2001nr,Cremades:2002dh,%
Kokorelis:2002zz,Kokorelis:2002qi}.
It was shown that chiral fermions indeed appear at the intersections
when the branes are given by 
(hyper)planes \cite{Steinacker:2013eya}, 
which can be
represented by operators or infinite dimensional matrices
in the matrix model.
The authors then proposed 
to replace these branes by fuzzy spheres and
other fuzzy manifolds, which can be represented by finite-$N$ matrices.
Realization of the Standard Model has also been discussed. 
%

Recently, two of the authors (J.N.\ and A.T.) \cite{Nishimura:2013moa}
calculated explicitly the spectrum of the Dirac operator
for a finite-$N$ configuration 
suggested in ref.~\cite{Chatzistavrakidis:2011gs},
which represents 
a 5-brane and a 7-brane intersecting at a point in the extra
dimensions.
It was confirmed that a chiral zero mode localized
at the intersection point indeed appears in the large-$N$ limit.
However, one also obtains
another chiral zero mode with opposite chirality,
which was not anticipated naively from the brane configuration.
This result was understood as a consequence of
a no-go theorem,
which states that
chiral fermions cannot be realized
in the large-$N$ limit of finite-$N$ 
type IIB matrix model
as far as one assumes
that space-time is given by
a direct product of our four-dimensional space-time and
the space in extra six dimensions.
In fact, the SO(3,1) Lorentz symmetry alone does not
imply the direct product structure of space-time, 
and one generally obtains a warp factor.
In a generic case with 
a nontrivial matrix $M$ representing a warp factor,
chiral zero modes in extra six dimensions do not automatically 
correspond to those in our four-dimensional space-time.
For the above explicit configuration,
it was found that there are huge degrees of freedom in $M$,
which allows only the desired chiral zero mode to appear
in four dimensions \cite{Nishimura:2013moa}.
Thus one can realize a chiral fermion in the large-$N$ limit
of finite-$N$ type IIB matrix model thanks to 
the matrix warp factor $M$.
%

The no-go theorem and the need for introducing a nontrivial $M$ 
to avoid its consequence affect drastically the discussions
on the possibilities of realizing chiral fermions and
the Standard Model 
in the type IIB matrix model.
In particular, the new insights enable us to realize chiral fermions
from intersecting
fuzzy ${\rm S}^2$ and fuzzy ${\rm S}^2 \times {\rm S}^2$, 
which can be obtained as classical solutions in the
type IIB matrix model assuming that a Myers term \cite{Myers:1999ps}
is induced dynamically.
(See refs.~\cite{Imai:2003vr,Imai:2003jb,Kaneko:2005pw},
which discuss the appearance of these fuzzy manifolds
in the type IIB matrix model due to quantum corrections.
Note also that,
including the dimensionality of our 4d space-time,
fuzzy ${\rm S}^2$ and fuzzy ${\rm S}^2 \times {\rm S}^2$
correspond to a D5-brane and a D7-brane, respectively,
which naturally appear in type IIB superstring theory.)
The two types of fuzzy manifold intersect in the six-dimensional space 
generically at even number of points, which give rise to
pairs of chiral fermions with opposite chirality in six dimensions.
However, by using the degrees of freedom in the matrix 
warp factor $M$, 
one can obtain only the desired chiral zero modes in four dimensions.

Extending this basic setup,
we discuss an explicit realization of the Standard Model.
The SU($n$) group can be realized as a subgroup of
U($n$), which appears naturally from $n$ coinciding branes.
First we introduce ``SU(3) branes'',
which consist of three coinciding fuzzy ${\rm S}^2 \times {\rm S}^2$,
and ``SU(2) branes'', 
which consist of two coinciding fuzzy ${\rm S}^2$.
In addition, we introduce a ``lepton brane'',
which is a single fuzzy ${\rm S}^2 \times {\rm S}^2$,
and an ``up-type brane'' and a ``down-type brane'', 
which are two separate fuzzy ${\rm S}^2$.
Thus we end up with a configuration 
with five stacks of branes intersecting with each other.
%
An important point 
here is that
chiral fermions actually appear only from intersections
of fuzzy ${\rm S}^2$ and fuzzy ${\rm S}^2 \times {\rm S}^2$.
This enables us to
obtain just the chiral fermions in the Standard Model
plus a right-handed neutrino, with the correct gauge interactions.
One can also check that the hypercharge can be assigned to
the chiral fermions consistently.

In fact, we show that the number of intersections of 
${\rm S}^2$ and ${\rm S}^2 \times {\rm S}^2$
in six dimensions cannot exceed four 
for arbitrary radii, location of the centers and their relative angles.
This implies that we can obtain only up to two generations
if we restrict ourselves to such configurations.
Three generations can be realized, for instance,
by squashing ${\rm S}^2$ or 
${\rm S}^2 \times {\rm S}^2$ that appear in the configuration.
We also discuss how the Higgs field appears from the bosonic matrices,
with nontrivial
Yukawa couplings to the Standard Model fermions.
Thus we find that all the peculiar features i)--iv) of the Standard
Model listed above can be explained from a rather simple structure
in the extra dimensions within the type IIB matrix model.
The main results of this paper was reported by A.T.\
at the Workshop on Noncommutative Field Theory and Gravity,
8-15 September 2013 held in Corfu, Greece.

The rest of this paper is organized as follows.
In section \ref{sec:chiral}
we briefly review how chiral fermions can be realized in 
the type IIB matrix model following ref.~\cite{Nishimura:2013moa}.
In section \ref{sec:fuzzyS2}
we discuss the emergence of chiral fermions from
a basic configuration, which consists of
fuzzy ${\rm S}^2$ and fuzzy ${\rm S}^2 \times {\rm S}^2$.
In section \ref{sec:SMfermions}
we discuss how one can realize the Standard Model fermions
by considering a matrix configuration corresponding
to five stacks of branes.
In section \ref{sec:num-generations}
we discuss the number of generations that can be realized
within this setup.
In section \ref{sec:Higgs} we discuss how the gauge field
and the Higgs field appear from the model.
In particular, we discuss how
nontrivial Yukawa couplings 
can be obtained from the overlap of wave functions.
Section \ref{sec:concl} is devoted to a summary and discussions.

~

\noindent {\bf Note Added:}
While we were preparing the manuscript, 
we encountered a preprint \cite{Steinacker:2014fja},
which has certain overlap with our paper.
%

\section{Realizing chiral fermions in the type IIB matrix model}
\label{sec:chiral}

The type IIB matrix model has an action \cite{IKKT}
\beqa
S &=& S_{\rm b} + S_{\rm f}  \ ,  
\label{ikkt-action}
\\
S_{{\rm b}} & = & 
-\frac{1}{4g^{2}}\mbox{Tr}\Bigl(
\left[A_{M},A_{N}\right]
\left[A^{M},A^{N}\right] \Bigr)
\ ,  
\label{S_b}
\\
S_{\rm f}&=&\frac{1}{2g^{2}}\, \mbox{Tr}
\Bigl(\bar{\Psi}\Gamma^M[A_M,\Psi] \Bigr) \ ,
\label{fermionic action}
\eeqa
%
where $\Gamma_M$ are $32 \times 32$ gamma matrices
in 10d.
The bosonic $N\times N$ matrices
$A_M$ ($M=0,\ldots,9$) are traceless Hermitian,
while
the fermionic $N\times N$ matrices 
$\Psi _\alpha$ ($\alpha = 1, \ldots , 32$)
are Majorana-Weyl fermions in 10d,
and, in particular, they satisfy
\begin{align}
\Gamma_{\chi} \Psi  &= \Psi \ ,
\label{Weyl-cond}
\end{align}
where $\Gamma_{\chi}$ is
the chirality operator in 10d.
Since the coupling constant $g$ can be absorbed 
by rescaling $A_{\mu}$ and $\Psi$ appropriately, it is merely
a scale parameter. 

The type IIB matrix model is conjectured to be a nonperturbative
definition of superstring theory \cite{IKKT}.
There are various pieces of evidence for this conjecture.
First of all, the action 
(\ref{ikkt-action})
can be regarded as a matrix regularization of
the worldsheet action of type IIB superstring theory
in the Schild gauge \cite{IKKT}.\footnote{This does not imply 
that the matrix model is merely a formulation
for the ``first quantization'' of superstrings.
In fact, multiple worldsheets appear naturally
in the matrix model as block-diagonal configurations,
where each block represents the embedding of a single worldsheet
into the 10-dimensional target space.}
Secondly, D-branes in type IIB superstring theory can be
described 
as simple matrix configurations,
and the interaction between
them can be reproduced correctly \cite{IKKT}.
Thirdly, 
under a few reasonable assumptions,
the string field Hamiltonian for type IIB superstring theory
can be derived from Schwinger-Dyson equations for the Wilson loop
operators, which are identified as creation and annihilation operators
of strings \cite{Fukuma:1997en}.

In all these connections to type IIB superstring theory, 
the target space coordinates are identified with the eigenvalues
of the matrices $A_\mu$ \cite{Aoki:1998vn}.
In particular, this identification is consistent with
the supersymmetry algebra of the model, in which the translation
that appears from the anti-commutator of supersymmetry generators
is identified with the shift symmetry
$A_\mu \mapsto A_\mu + \alpha_\mu {\bf 1}$
of the model, where $\alpha_\mu \in {\bf R}$.
Also the fact that the model has extended ${\cal N}=2$ supersymmetry
in ten dimensions is consistent with the assertion 
that the model actually includes gravity since it is known in field theory
that ${\cal N}=1$ supersymmetry is the maximal one 
that can be achieved in ten dimensions
without including gravity.

Below we review the general arguments 
in ref.~\cite{Nishimura:2013moa}
concerning the appearance
of chiral fermions in 4d from the type IIB matrix model.
In this model, the space-time is represented
by the ten bosonic $N \times N$ Hermitian matrices $A_M$
($M = 0 , \ldots , 9$).
As it was shown by ref.~\cite{Kim:2011cr}, an
expanding three-dimensional space appears dynamically
after some time.
At later times, it is speculated that three-dimensional space
becomes much larger than 
the typical scale of the model, and that
quantum fluctuations can be neglected 
at large scales \cite{Kim:2011ts,Kim:2012mw}.
Furthermore, as far as we do not consider too long time scale,
we can neglect the expansion of space and therefore the space-time
has SO(3,1) 
Lorentz symmetry.
Thus we are led to consider matrix configurations given by
\begin{align}
A_\mu &= X_\mu \otimes M  \quad (\mu=0, \ldots ,3) \ ,
\label{Amu}
\\
A_a &= \1_n \otimes  Y_a  \quad (a=4, \ldots ,9) \ .
\label{Aa}
\end{align}
Here we assume that the $n \times n$ Hermitian matrices
$X_\mu$ have the property
$O_{\mu\nu} X_\nu = g[O] \, X_\mu \, g[O]^\dag$,
where $O \in {\rm SO}(3,1)$ and $g[O]\in {\rm SU}(n)$.
Then (\ref{Amu}) and (\ref{Aa}) can be
regarded as the most general configuration that is
SO(3,1) 
invariant up to ${\rm SU}(N)$ symmetry.

The Hermitian matrix $M$ in (\ref{Amu})
can be regarded as a matrix version
of the warp factor. The special case $M=\1$ corresponds
to a space-time which is a direct product of (3+1)-dimensional
space-time and the extra dimensions.
However, from the viewpoint of 
preserving the Lorentz symmetry, there is no reason
to set $M=\1$.
%


In order to discuss chiral fermions in 4d, 
it is convenient to decompose the gamma matrices in 10d 
into the ones in 4d and 6d as
\begin{align}
\Gamma^{\mu}&=\gamma^{\mu}\otimes \1_8 \ , \nonumber\\
\Gamma^a&= i \gamma_{\chi}^{\rm (4d)} \otimes \Delta^a \ ,
\label{gamma matrices}
\end{align}
where 
$\gamma^{\mu}$ and $\Delta^a$ are gamma matrices in 4d and 6d, 
respectively, which satisfy
\begin{align}
\{\gamma^{\mu},\gamma^{\nu}\}&=-2\eta^{\mu\nu} \ , \nonumber\\
\{\Delta^a,\Delta^b\}&=2\delta^{ab} \ ,
\end{align}
and $\gamma_{\chi}^{(4d)}$ is the chirality operator in 4d.
Note that the chirality operator
$\Gamma_{\chi}$ in 10d can be decomposed as
\begin{align}
\Gamma_{\chi} = \gamma_{\chi}^{\rm (4d)} \otimes
\Delta_{\chi}^{\rm (6d)} \ ,
\label{chi-decompose}
\end{align}
where $\Delta_{\chi}^{\rm (6d)}$ is
the chirality operator in 6d.

In the case of quantum field theory in higher dimensions,
one decomposes fields into
Kaluza-Klein modes, which can then be identified
as four-dimensional fields.
Here we make a similar analysis in the language of 
matrices.\footnote{See ref.~\cite{Nishimura:2012rs} 
for discussions on the appearance of local field theory
in the Lorentzian type IIB matrix model.}
We consider expanding the fermionic variables
in terms of the eigenmodes of
the Dirac operator in 6d defined by
\begin{align}
D_{\rm 6d}\Phi=\Delta^a[Y_a,\Phi] \ .
\label{Dirac operator in 6d}
\end{align}
In the explicit example to be discussed in the next section,
we consider a configuration of $Y_a$, which has a block diagonal form
\begin{align}
Y_a= \left( 
      \begin{array}{cc}
       Y_a^{(1)} & 0 \\
       0   & Y_a^{(2)} 
      \end{array} \right) \ .
\label{Ya-block}
\end{align}
Correspondingly, we decompose $\Phi$ 
in eq.~(\ref{Dirac operator in 6d}) as
\begin{align}
\Phi= \left( 
      \begin{array}{cc}
       \Phi^{(1,1)} & \Phi^{(1,2)} \\
       \Phi^{(2,1)}   & \Phi^{(2,2)}
      \end{array} \right) \ .
\label{decomposition into blocks}
\end{align}
Since the Dirac operator $D_{\rm 6d}$ acts 
on each block $\Phi^{(I,J)} \;(I,J=1,2)$ independently,
the eigenvalue problem for $D_{\rm 6d}$ can be decomposed
into that in each block.

As we will see in the explicit example,
chiral fermions actually appear in off-diagonal blocks.
Therefore, from now on,
we consider the eigenvalue problem for 
$\varphi \equiv \Phi^{(1,2)}$, which is given by
\begin{align}
\Delta^a(Y_a^{(1)}\varphi-\varphi Y_a^{(2)})=\lambda \, \varphi \ .
\label{eigenvalue equation}
\end{align}
Due to the fact that $Y_a$ and $\Delta^a$ are Hermitian matrices, 
one can easily show that the eigenvalue $\lambda$ in 
(\ref{eigenvalue equation}) is real.
Also, by multiplying $\Delta_{\chi}^{\rm (6d)}$ to
(\ref{eigenvalue equation}) from the left, one obtains
\begin{align}
\Delta^a \Bigl\{ Y_a^{(1)}( \Delta_{\chi}^{\rm (6d)}\varphi)
-(\Delta_{\chi}^{\rm (6d)}\varphi) Y_a^{(2)} \Bigr\}
=-\lambda \, (\Delta_{\chi}^{\rm (6d)}\varphi) \ .
\label{eigenvalue equation 2}
\end{align}
This implies that 
if $\varphi$ is an eigenvector with 
the eigenvalue $\lambda$, 
$\Delta_{\chi}^{\rm (6d)}\varphi$
is an eigenvector with the eigenvalue $-\lambda$.
In particular, $\varphi$ and $\Delta_{\chi}^{\rm (6d)}\varphi$ 
are linearly independent for $\lambda \neq 0$.
Therefore we can construct left-handed and right-handed modes 
by taking linear combinations of
$\varphi$ and $\Delta_{\chi}^{\rm (6d)}\varphi$ 
with $\lambda \neq 0$ as
\begin{align}
\varphi_R &=\frac{1+\Delta_{\chi}^{\rm (6d)}}{2}\varphi \ , 
\nonumber\\
\varphi_L &=\frac{1-\Delta_{\chi}^{\rm (6d)}}{2}\varphi \ ,
\label{phi-LRdecomposed}
\end{align}
which satisfy
\begin{align}
\Delta_{\chi}^{\rm (6d)}\varphi_R&=\varphi_R  \ , \nonumber\\
\Delta_{\chi}^{\rm (6d)}\varphi_L&=-\varphi_L  \ , 
\label{6d chirality}
\end{align}
and
\begin{align}
\Delta^a(Y_a^{(1)}\varphi_R-\varphi_R Y_a^{(2)})
&=\lambda \, \varphi_L \ , \nonumber\\
\Delta^a(Y_a^{(1)}\varphi_L-\varphi_L Y_a^{(2)})
&=\lambda \, \varphi_R \ .
\label{6d Dirac equation}
\end{align}
Thus the non-zero modes appear in pairs of right-handed and left-handed modes.
On the other hand,
the zero modes can be 
assumed to have definite chirality.
Since we are considering finite-$N$ matrices,
the space of $\varphi$ with each chirality
has the same dimension.
Therefore, the number of zero modes with each chirality 
should also be the same.
However, the actual form of the zero mode $\varphi$ with each chirality
can be very different in general. 
This fact will be important in getting chiral fermions in 4d.

Let $\{\lambda_n\}$ be a set of non-negative eigenvalues 
in (\ref{eigenvalue equation}).
Then we denote the right-handed and left-handed modes 
corresponding to $\lambda_n$ 
by $\varphi_{nR}$ and $\varphi_{nL}$, respectively.
These modes can be normalized in such a way that 
they satisfy the orthonormal condition
\begin{align}
\mbox{tr}(\varphi_{m A}^{\dagger}\varphi_{n B})=\delta_{mn}\delta_{AB} \ ,
\label{orthonormality condition}
\end{align}
where the labels $A$ and $B$ represent either $R$ or $L$.

Now we decompose the fermionic variables $\Psi$ in 
(\ref{fermionic action})
in the same way as in (\ref{decomposition into blocks}), and 
expand the off-diagonal block $\Psi^{(1,2)}$ 
in terms of the orthonormal basis 
$\varphi_{nR}$ and $\varphi_{nL}$ constructed above as
\begin{align}
\Psi^{(1,2)}=\sum_n (\psi_{nR}\otimes \varphi_{nR} 
+ \psi_{nL}\otimes \varphi_{nL} ) \ .
\label{expansion of Psi}
\end{align}
Note that the matrix coefficients $\psi_{nR}$ and $\psi_{nL}$
introduced here satisfy
\begin{align}
\gamma_{\chi}^{\rm (4d)} \psi_{nR} &= \psi_{nR}  \ , \nonumber\\
\gamma_{\chi}^{\rm (4d)} \psi_{nL} &= -\psi_{nL} \ ,
\label{4d chirality}
\end{align}
as one can see from (\ref{Weyl-cond}), (\ref{chi-decompose}) 
and (\ref{6d chirality}).
Namely, the left-handed and right-handed modes 
in 6d correspond to the left-handed and right-handed modes 
in 4d, respectively.
Note also that the other off-diagonal block $\Psi^{(2,1)}$
is related to the off-diagonal block $\Psi^{(1,2)}$ that
we have considered through charge conjugation as
\begin{align}
\Bigl(\Psi^{(2,1)}\Bigr)^{\rm c}=\Psi^{(1,2)}  \ ,
\label{Majorana condition}
\end{align}
due to the Majorana condition for $\Psi$.
This allows us to focus only on $\Psi^{(1,2)}$.

In what follows we consider the case in which
the matrix warp factor $M$ in (\ref{Amu}) has a block diagonal form
similar to (\ref{Ya-block}) as
\begin{align}
M =\left( \begin{array}{cc}
                M^{(1)} & 0 \\
                0 & M^{(2)}
        \end{array} \right) \ .
\label{block diagonal M}
\end{align}
Then, by substituting (\ref{expansion of Psi}) into the action 
(\ref{fermionic action}), 
we find that the case $M=\1$, namely 
the case with 
$M^{(1)}=\1$ and $M^{(2)}=\1$ in (\ref{block diagonal M}),
leads to vector-like fermions in 4d \cite{Nishimura:2013moa}.
(See also footnote \ref{foot:vector-like}.)
Therefore we find that in order to obtain chiral fermions,
we need to consider $M \neq \1$ and nonvanishing $Y_a$.
In particular, let us consider the case with 
$\lambda_0=0$ and $\lambda_n \neq 0$ for $n \neq 0$.
Suppose $M^{(1)}$ and $M^{(2)}$ satisfy
\begin{align}
M^{(1)}\varphi_{0A}=\varphi_{0A}M^{(2)}=\varphi_{0A} \ ,
\label{condition for M}
\end{align}
for the left-handed mode $A=L$, but not for the right-handed mode $A=R$ .
Then, we find that
$\psi_{0L}$ has an appropriate form of the action as 
a chiral fermion in 4d, and
it does not couple to the other modes \cite{Nishimura:2013moa}. 
On the other hand, $\psi_{0R}$ does not have
an appropriate form of the action as 
a chiral fermion in 4d, and
it couples to the massive modes with $\lambda_n \neq 0$.
This implies that we obtain only a left-handed chiral fermion in 4d.
While we do not know at present the
mechanism that favors the matrix $M$
satisfying (\ref{condition for M}) for only one of the chiralities,
we will argue that all the chiral fermions in the Standard Model
can be obtained in this way
by using the huge degrees of freedom in $M$.
%


\section{A basic configuration with fuzzy ${\rm S}^2$ and 
fuzzy ${\rm S}^2 \times {\rm S}^2$}
\label{sec:fuzzyS2}

As an example of $Y_a$, which allows 
nontrivial solutions to (\ref{eigenvalue equation}) 
with $\lambda=0$, 
we consider an explicit finite-$N$ configuration
%
given by
\begin{align}
Y_4 &= \frac{1}{r} \left(
\begin{array}{cc}
L_1 & 0 \\
0   & \1_k \otimes \tilde{L}_3
\end{array} \right) \ ,
&
Y_5 &= \frac{1}{r} \left(
\begin{array}{cc}
L_2 & 0 \\
0   & \tilde{L}_3 \otimes \1_k
\end{array} \right) \ ,
&
Y_6 &= \frac{1}{r} \left(
\begin{array}{cc}
\tilde{L}_3 & 0 \\
0   & L_1 \otimes \1_k
\end{array} \right) \ ,
\nonumber \\
Y_7 &= \frac{1}{r} \left(
\begin{array}{cc}
0  & 0 \\
0   & L_2 \otimes \1_k
\end{array} \right) \ ,
&
Y_8 &=\frac{1}{r} \left(
\begin{array}{cc}
0   & 0 \\
0   & \1_k \otimes L_1
\end{array} \right) \ ,
&
Y_9 &= \frac{1}{r} \left(
\begin{array}{cc}
0   & 0 \\
0   & \1_k \otimes L_2
\end{array} \right) \ ,
\label{S2-S2S2-config}
\end{align}
where we have defined
\begin{align}
\tilde{L}_3\equiv  L_3+ r \1_k \ .
\label{shifting-L3}
\end{align}
Here, the $k\times k$ matrices $L_i$ ($i=1,2,3$)
are 
the $k$-dimensional irreducible representation of 
the SU(2) algebra $[L_i , L_j ] = i \epsilon_{ijk} L_k$,
which represents a fuzzy sphere
\begin{align}
\sum_{i=1}^3 (L_i)^2 = r^2  \1  
\end{align}
with the radius $r=\frac{1}{2} \sqrt{k^2 -1}$.
The top-left block and the bottom-right block 
in eq.~(\ref{S2-S2S2-config})
represent fuzzy ${\rm S}^2$ and
fuzzy ${\rm S}^2 \times {\rm S}^2$, respectively, 
and eq.~(\ref{shifting-L3}) amounts to shifting them
in some directions.
Including the dimensionality in four-dimensional space-time,
these fuzzy manifolds correspond to a D5-brane and a D7-brane, 
respectively, which appear naturally
in type IIB superstring theory.

\begin{figure}[t]
  \begin{center}
  \includegraphics[height=5.5cm]{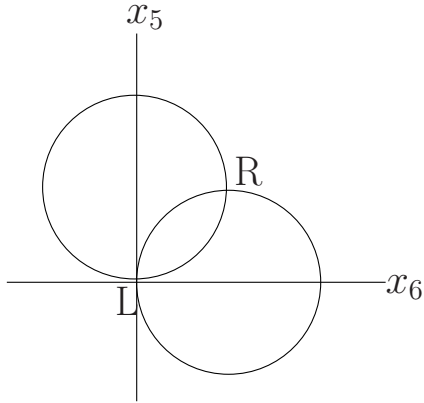}
\end{center}
  \caption{The intersections of the
${\rm S}^2$ and the ${\rm S}^2 \times {\rm S}^2$,
which are given by
(\ref{S2-eq}) and (\ref{S2S2-eq}), respectively.
We have two intersecting points, at which chiral zero modes
appear with the chirality shown by L and R for 
left-handed and right-handed
fermions, respectively.}
  \label{figure:one-gen}
  \end{figure}

The intersections of the two fuzzy manifolds
in eq.~(\ref{S2-S2S2-config})
can be obtained easily.  
The classical manifold corresponding to the fuzzy ${\rm S}^2$ is 
represented by
\begin{align}
(x_4)^2 + (x_5)^2 + (x_6 - 1 )^2  = 1  \quad
\mbox{and} \quad x_7 = x_8 = x_9 = 0\ ,
\label{S2-eq}
\end{align}
whereas that
corresponding to 
the fuzzy ${\rm S}^2 \times {\rm S}^2$ is represented by
\begin{align}
(x_5 - 1 )^2 + (x_6)^2 + (x_7)^2 = 1  \quad
\mbox{and} \quad 
(x_4 - 1 )^2 + (x_8)^2 + (x_9)^2 = 1 \ .
\label{S2S2-eq}
\end{align}
Solving (\ref{S2-eq}) and (\ref{S2S2-eq}) simultaneously, 
we obtain $(x_5,x_6)=(0,0)$ and $(1,1)$ with $x_4=x_7=x_8=x_9=0$,
which represent two intersecting points in the 6d space.
At the two intersecting points,
we obtain chiral fermions with opposite chirality.
The situation
is depicted on the $(x_5,x_6)$-plane
in figure \ref{figure:one-gen}.

The configuration (\ref{S2-S2S2-config})
looks similar to the one studied in ref.~\cite{Nishimura:2013moa}
following the original proposal \cite{Chatzistavrakidis:2011gs}.
However, there are some important differences
in the bottom-right block.
The $Y_8$ in eq.~(17) of ref.~\cite{Nishimura:2013moa}
involves $\tilde{L}_3 \otimes L_1$.
This implies that the bottom-right block 
in ref.~\cite{Nishimura:2013moa}
does not represent a fuzzy ${\rm S}^2 \times {\rm S}^2$,
and hence it cannot be obtained as a classical solution
in a matrix model with a Myers term.
Moreover, the extent of the configuration 
in ref.~\cite{Nishimura:2013moa}
diverges in the $x_8$-direction 
when one takes the large-$k$ limit.
On the other hand, the configuration
(\ref{S2-S2S2-config}) is completely finite 
in the large-$k$ limit. This is more natural
in view of the results obtained
by Monte Carlo simulation \cite{Kim:2011cr},
which show that the extent in the six dimensions remains finite and
small.

\begin{figure}[t]
\begin{center}
\includegraphics[height=5.5cm]{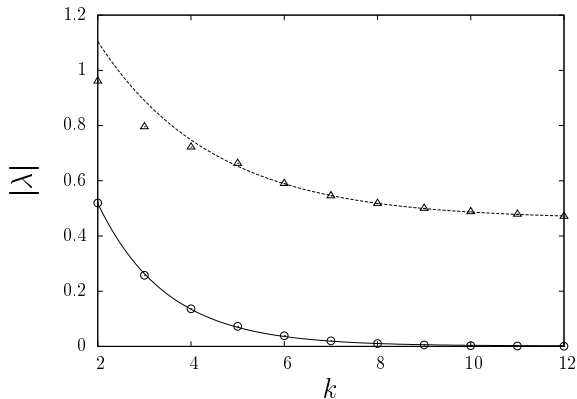}
\end{center}
\caption{
The smallest $|\lambda|$ (circles) and the second smallest one (triangles)
are plotted against $k$
for the basic configuration (\ref{S2-S2S2-config}).
The solid and dashed lines 
represent the fits to the behavior 
$|\lambda| = a + b \, {\rm exp}(- c \, k)$.
For the smallest $|\lambda|$, we find 
$a=0.0016(13)$, $b=2.00(4)$, $c=0.68(1)$
using the fitting range $2 \le k \le 12$. 
The obtained value of $a$ is consistent
with zero within the fitting error.
For the second smallest one, we obtain
$a=0.461(2)$, $b=1.4(1)$, $c=0.40(2)$
using the fitting range $6 \le k \le 12$. 
}
\label{fig:lambda-k}
\end{figure}


We would like to solve 
the eigenvalue problem
for the Dirac operator in 6d defined
by (\ref{Dirac operator in 6d})
with $Y_a$ given above.
Due to the block diagonal structure of $Y_a$, 
we can decompose $\Phi$ into blocks as
(\ref{decomposition into blocks}),
and the problem reduces
to solving the equation for each block.
On physical grounds,
it is expected that chiral zero modes appear from the 
off-diagonal blocks, which correspond to the degrees of freedom
connecting the two different branes.
Thus the problem reduces to
solving the eigenvalue equation
(\ref{eigenvalue equation}).
As we explained below eq.~(\ref{eigenvalue equation 2}),
eigenvalues $\lambda$
appear in pairs with opposite signs.
%
In figure~\ref{fig:lambda-k} the smallest $|\lambda|$
and the second smallest one are plotted against $k$.
We find that the smallest one vanishes rapidly with increasing $k$,
which implies the appearance of two zero modes in the large-$k$ limit.
The second smallest $|\lambda|$, on the other hand, 
seems to approach a non-zero constant.

\begin{figure}[t]
\begin{center}
\includegraphics[width=72mm]{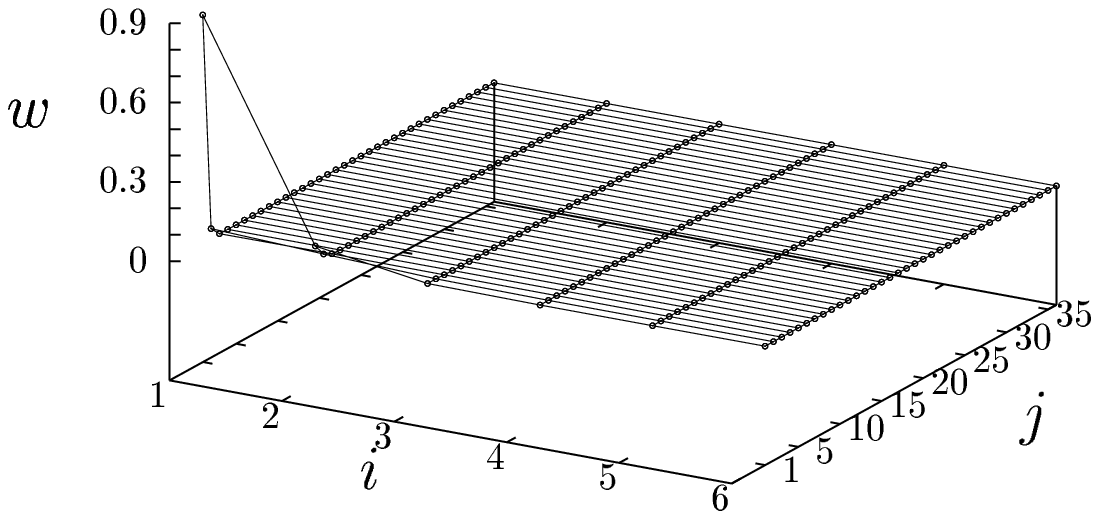}
\includegraphics[width=72mm]{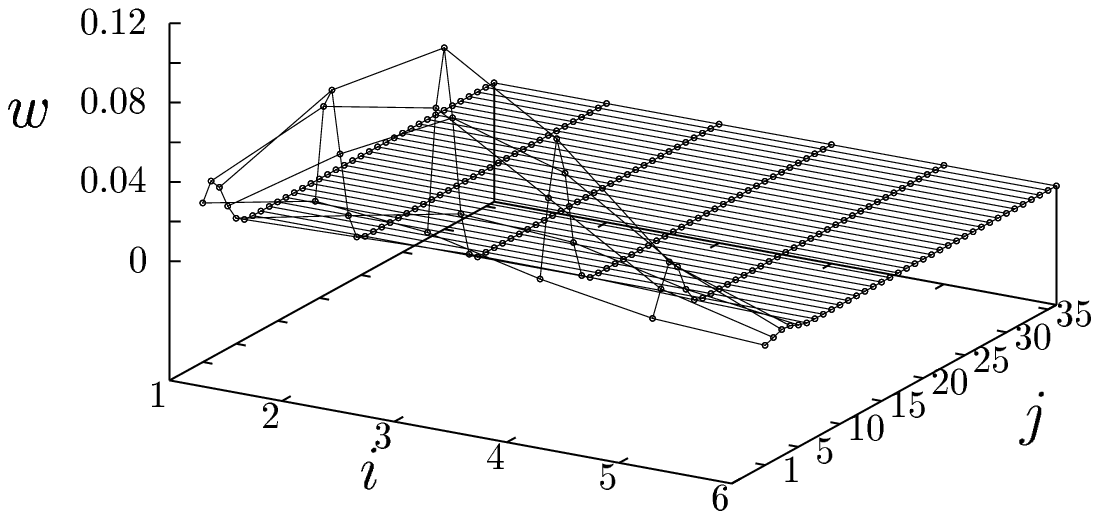}
\end{center}
\caption{
The shape of the wave functions $w(i,j)$ 
of the chiral zero modes with each chirality
found in figure \ref{fig:lambda-k} is plotted for $k=6$.
(Left) $w(i,j)$ for the left-handed mode.
A clear peak is seen at $(i,j)=(1,1)$.
(Right) $w(i,j)$ for the right-handed mode.
It is widely spread within the region 
$1\le i \le k$ and $1\le j \le k$.
However, by change of the basis,
one can make the right-handed mode localized, whereas the
left-handed mode is not. See figure 4.
}
\label{fig:wf}
\end{figure}

Let us take an appropriate linear combination of the two modes
with the smallest $|\lambda|$ so that they have definite chirality
as we have done in eq.~(\ref{phi-LRdecomposed}).
Figure~\ref{fig:wf} shows the shape of the wave function
\begin{align}
w(i,j) \equiv \sum_{\alpha=1}^8 |(\varphi_\alpha)_{ij}|^2 
\quad \quad
(1 \le i \le k \ , \quad 1 \le j \le k^2)
\label{shape-wf}
\end{align}
of the chiral mode with each chirality for $k=6$.
Let us recall here that
$(\varphi_\alpha)_{ij}$ represents the top-right block 
$\Phi^{(1,2)}$ in (\ref{decomposition into blocks}).
In the eigenvalue equation (\ref{eigenvalue equation}),
the $k\times k$ matrices $Y_a^{(1)}$ act on
the index $i$, whereas  
the $k^2 \times k^2$ matrices $Y_a^{(2)}$ act on
the index $j$. In the present configuration (\ref{S2-S2S2-config}),
$Y_a^{(1)}$ and $Y_a^{(2)}$ represent the fuzzy ${\rm S}^2$ 
and the fuzzy ${\rm S}^2 \times {\rm S}^2$, respectively.
Below we explain some conventions used to make the plots 
in figure~\ref{fig:wf}.
First, as a $k$-dimensional representation
of the ${\rm SU}(2)$ algebra 
in the configuration (\ref{S2-S2S2-config}), we use
\begin{align}
(L_1)_{mn}&=\frac{1}{2}\sqrt{n(k-n)}
\delta_{m,n+1}+\frac{1}{2}\sqrt{(n-1)(k-n+1)}\delta_{m,n-1} \ , \nonumber\\
(L_2)_{mn}&=\frac{1}{2i}\sqrt{n(k-n)}\delta_{m,n+1}
-\frac{1}{2i}\sqrt{(n-1)(k-n+1)}\delta_{m,n-1} \ , \nonumber\\
(L_3)_{mn}&= \left( n-\frac{k+1}{2} \right)\delta_{mn} \ .
\label{matrix-elements-for-SU2}
\end{align}
Second, when we make a tensor product 
$Y_a^{(2)} = A_a \otimes B_a$
in the bottom-right block in (\ref{S2-S2S2-config}),
we combine the 
indices as $(Y_a^{(2)})_{j j'}=(A_a)_{pq} (B_a)_{rs}$ 
with $j = k(r-1)+p$ and $j' = k(s-1)+q$.
In this way, the indices $i$ and $j$ of the wave function 
$(\varphi_\alpha)_{ij}$ correspond to the eigenvalues of
$L_3$ in the fuzzy ${\rm S}^2$ 
and the fuzzy ${\rm S}^2 \times {\rm S}^2$, respectively.

From figure \ref{fig:wf} (Left)
we find that the left-handed chiral mode
has 
a peak at $(i,j)=(1,1)$.
With the chosen conventions,
$i=1$ corresponds
to the point $(x_4,x_5,x_6)=(0,0,0)$ 
on the fuzzy ${\rm S}^2$ (\ref{S2-eq}),
whereas $j=1$ corresponds
to the point $(x_5,x_6,x_7)=(0,0,0)$ and
$(x_4,x_8,x_9)=(0,0,0)$
on the fuzzy ${\rm S}^2 \times {\rm S}^2$ (\ref{S2S2-eq}).
Thus we find that the wave function of the left-handed
chiral mode is localized at one of the intersection points
$(x_5,x_6)=(0,0)$ in figure \ref{figure:one-gen}.

\begin{figure}[t]
\begin{center}
\includegraphics[width=72mm]{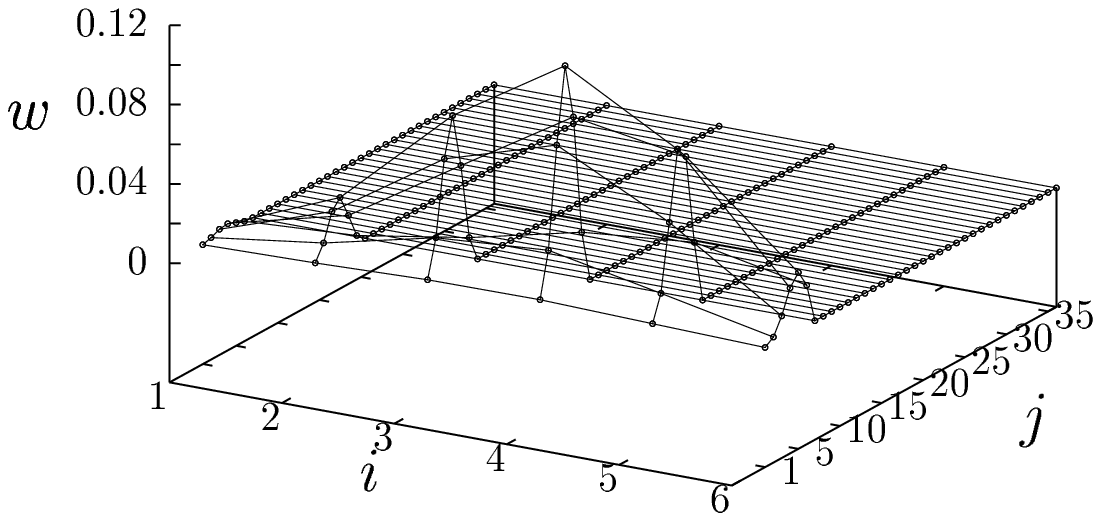}
\includegraphics[width=72mm]{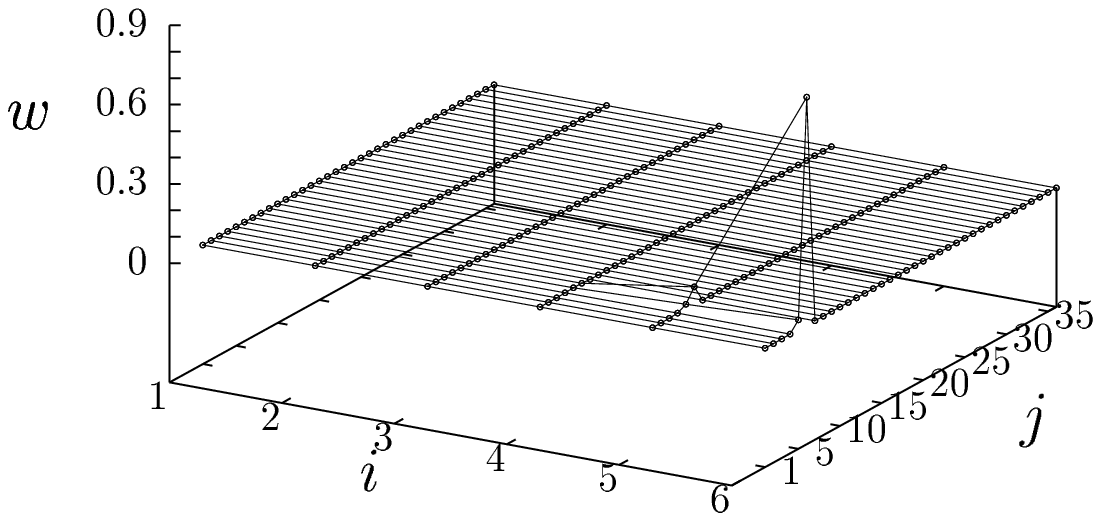}
\end{center}
\caption{
The shape of the wave functions $w(i,j)$ 
of the chiral zero modes with each chirality
found in figure \ref{fig:lambda-k} is plotted for $k=6$.
We make a change of the basis (\ref{change-S2-basis})
and (\ref{change-S2S2-basis}).
(Left) $w(i,j)$ for the left-handed mode.
It is widely spread within the region 
$1\le i \le k$ and $1\le j \le k$.
(Right) $w(i,j)$ for the right-handed mode.
A clear peak is seen at $(i,j)=(6,6)$.
}
\label{fig:wf-prm}
\end{figure}

On the other hand, 
from figure \ref{fig:wf} (Right),
we find that the wave function of the right-handed
chiral mode does not seem to be 
localized.\footnote{In fact, 
the wave function is suppressed at $j>k$, which
implies that it is localized
at $(x_4,x_8,x_9)=(0,0,0)$
on the second ${\rm S}^2$ of ${\rm S}^2 \times {\rm S}^2$ (\ref{S2S2-eq}).
}
However, this is simply due to the chosen representation
of the SU(2) algebra.
For instance,
let us make a replacement
\begin{align}
L_2 \mapsto L_3 \ , \quad L_3 \mapsto - L_2 
\label{change-S2-basis}
\end{align}
in the fuzzy ${\rm S}^2$, and a replacement
\begin{align}
\label{change-S2S2-basis}
L_1 \mapsto L_3 \ , \quad L_3 \mapsto - L_1 
\end{align}
in the first ${\rm S}^2$ of the fuzzy ${\rm S}^2 \times {\rm S}^2$.
The shape of the wave function (\ref{shape-wf}) with
this representation is shown in figure \ref{fig:wf-prm}.
We find that the right-handed chiral mode
has a peak at $(i,j)=(6,6)$.
With the new convention,
$i=6$ corresponds to the point $(x_4,x_5,x_6)=(0,1,1)$ 
on the fuzzy ${\rm S}^2$ (\ref{S2-eq}),
whereas $j=6$ corresponds
to the point $(x_5,x_6,x_7)=(1,1,0)$ and
$(x_4,x_8,x_9)=(0,0,0)$
on the fuzzy ${\rm S}^2 \times {\rm S}^2$ (\ref{S2S2-eq}).
This implies that the wave function of the right-handed
chiral mode is localized at one of the intersection points
$(x_5,x_6)=(1,1)$ in figure \ref{figure:one-gen}.
Thus, in order to see that chiral modes
are localized on the intersection points
from the profile of the wave function,
one generally needs to choose the representation of the
SU(2) algebra for each chiral mode appropriately.

In the original representation 
(\ref{S2-S2S2-config}) with (\ref{matrix-elements-for-SU2}),
the left-handed chiral mode takes a simple form 
\begin{align}
\varphi_{L\alpha}=
\left(
\begin{array}{ccccc}
\chi_{\alpha} & 0      & \cdots & \cdots  & 0      \\
0             & 0      & \cdots & \cdots  & 0      \\
\vdots        & \vdots &        &         & \vdots \\
0             & 0      & \cdots & \cdots  & 0    
\end{array}
  \right)   
\label{phiL-wf}
\end{align}
at large $k$, where $\sum_{\alpha}|\chi_{\alpha}|^2=1$.
Note that (\ref{phiL-wf}) is a $k \times k^2$ matrix.
The matrices $M^{(1)}$ and $M^{(2)}$ which satisfy 
(\ref{condition for M}) for this mode in the large-$k$ limit
are given by
\begin{align}
M^{(1)} &=\left(
\begin{array}{cccc}
1      & 0        & \cdots & 0 \\
0      & *        & \cdots & * \\
\vdots & \vdots   &        & \vdots  \\
0      & *        & \cdots & *
\end{array} \right) \ , \quad \quad 
M^{(2)} =\left(
\begin{array}{ccccc}
1      & 0       & \cdots & \cdots & 0       \\
0      & *       & \cdots & \cdots & *       \\
\vdots & \vdots  &        &        & \vdots  \\
\vdots & \vdots  &        &        & \vdots  \\
0      & *       & \cdots & \cdots & *
\end{array} \right) \ ,
\label{M-example}
\end{align}
which are $k\times k$ and $k^2\times k^2$ 
Hermitian matrices, respectively.
In the same representation, 
the wave function of
the right-handed chiral mode is quite different 
from (\ref{phiL-wf}), and it does not satisfy
(\ref{condition for M}) generically.
Thus we can obtain a single chiral zero mode in 
four dimensions.




\begin{figure}[t]
\begin{center}
  \includegraphics[height=5.5cm]{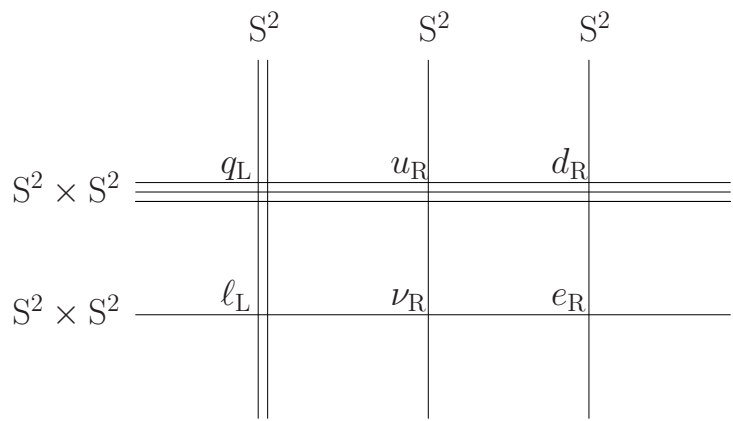}
\end{center}
\caption{A schematic view of the configuration with
five stacks of branes, which gives rise to 
the Standard Model fermions and a right-handed neutrino.}
\label{figure:SM-branes}
\end{figure}

\section{Realizing the Standard Model fermions}
\label{sec:SMfermions}

In this section we discuss how to realize the Standard Model
fermions extending the basic setup in section \ref{sec:fuzzyS2}.

First we introduce gauge symmetry 
by replacing each of the branes by coincident multiple branes
similarly to the case of D-brane effective theory.
For instance, if we make a replacement
\begin{align}
Y_a ^{(1)}  \mapsto Y_a ^{(1)} \otimes \1_{p} \ , \quad
Y_a ^{(2)}  \mapsto Y_a ^{(2)} \otimes \1_{q}  
\label{multiple-branes}
\end{align}
in the configuration (\ref{Ya-block}),
we obtain ${\rm U}(p) \times {\rm U}(q)$ gauge symmetry
as a subgroup of the U($N$) symmetry of the original model.
Then the chiral zero mode that appears from 
the top-right block 
$\Phi^{(1,2)}$ in (\ref{decomposition into blocks})
becomes a bi-fundamental representation $(p , \bar{q})$.

In order to realize the Standard Model fermions, 
we consider a matrix configuration with five diagonal blocks, 
$Y^{(1)}_a \otimes \1 _{p_1} ,
\ldots,
Y^{(5)}_a \otimes \1 _{p_5}$,
which correspond to five stacks of branes. 
First we introduce ``SU(3) branes'',
which consist of three coinciding fuzzy ${\rm S}^2 \times {\rm S}^2$,
and ``SU(2) branes'', 
which consist of two coinciding fuzzy ${\rm S}^2$.
In addition, we introduce a ``lepton brane'',
which is a single fuzzy ${\rm S}^2\times {\rm S}^2$,
and an ``up-type brane'' and a ``down-type brane'',
which are two separate fuzzy ${\rm S}^2 $.
Thus we end up with a configuration with five stacks of
branes\footnote{An analogous configuration was discussed
in section 4.2 of ref.~\cite{Chatzistavrakidis:2011gs}, 
but it was dismissed for a reason that does not apply to
our case.} depicted in figure \ref{figure:SM-branes}.




\begin{table}
\begin{center}
\begin{tabular}{c||c|c|c|c|c|c}
&$q$&$u$&$d$&$l$&$\nu$&$e$ \\ \hline\hline
$Y$&1/6&2/3&-1/3&-1/2&0&-1 \\ \hline
$B$&1/3&1/3&1/3&0&0&0 \\ \hline
$L$&0&0&0&1&1&1 \\ \hline
$Q_L$&1&0&0&1&0&0 \\ \hline
$Q_R$&0&1&1&0&1&1 
\end{tabular} 
\hspace{5mm}
\begin{tabular}{c||c|c|c|c|c}
&$c^1$&$c^2$&$c^3$&$c^4$&$c^5$ \\ \hline\hline
$Y$&1/6&-1/2&0&-1/2&1/2 \\ \hline
$B$&1/3&0&0&0&0 \\ \hline
$L$&0&1&0&0&0 \\ \hline
$Q_L$&0&0&-1&0&0 \\ \hline
$Q_R$&0&0&0&-1&-1 
\end{tabular} 
\caption{(Left) The hypercharge $Y$, the baryon number $B$, 
the lepton number $L$,
the left-handed charge $Q_L$ and the right-handed charge $Q_R$ 
for each chiral fermion are shown.
(Right) The coefficients $c^i$ in (\ref{lccQ}) 
for each kind of charge are shown.
The labels $i=1,\ldots,5$ correspond to
the SU(3) branes, the lepton brane, the SU(2) branes, 
the up-type brane and the down-type brane, respectively.}
\label{table:U(1)charges}
\end{center}
\end{table}

In fact, chiral fermions appear only from intersections
of fuzzy ${\rm S}^2$ and fuzzy ${\rm S}^2 \times {\rm S}^2$
in generic situations in six dimensions.
Note first that 
${\rm S}^2$ and ${\rm S}^2 \times {\rm S}^2$ 
have six dimensions in total,
and therefore they intersect with each other 
at some points in six dimensions generically.
On similar grounds,
two ${\rm S}^2$ do not intersect with each other.
Two ${\rm S}^2 \times {\rm S}^2$ intersect, but
on a two-dimensional surface.
In this case, the index in six dimensions vanishes
unless a nontrivial flux is induced on the surface.
Thus, it is possible to 
obtain just the chiral fermions in the Standard Model
plus a right-handed neutrino, with the correct gauge 
interactions.\footnote{For this purpose alone,
one may exchange the roles of ${\rm S}^2$ and
${\rm S}^2 \times {\rm S}^2$ simultaneously.
We prefer the present version, which makes our discussion
on the Higgs field simpler.
}
One can easily see that the chiral fermions have the correct representations
under the gauge group SU(3) $\times$ SU(2).

The hypercharge can also be assigned to the fermions correctly.
Each stack of branes produces a U(1) gauge group.
Let $Q^i$ ($i=1,\ldots,5$) be the U(1) charge associated with the $i$-th
stack of branes.
An open string connecting the $i$-th and $j$-th stacks of branes,
which is represented by 
the off-diagonal block connecting the $i$-th and $j$-th diagonal blocks,
has $Q^i = 1$ and $Q^j = -1$.
Therefore, if we define a charge ${\cal Q}$ as the linear combination
\beq
{\cal Q}= \sum_{i=1}^5 c^i Q^i \ ,
\label{lccQ}
\eeq
the chiral fermion that appears at the intersection of 
the $i$-th and $j$-th stacks of branes has ${\cal Q}=c^i -c^j$.
By appropriately choosing the coefficients $c^i$,
one can assign the correct hypercharge 
to the chiral fermions appearing at all the intersections.
Other choices of the coefficients $c^i$ define other U(1) charges
such as the baryon number $B$, the lepton number $L$,
the left-handed charge $Q_L$ and the right-handed charge $Q_R$.
The charges for each chiral fermion and 
the values of the coefficients $c^i$ for each charge
are given in table \ref{table:U(1)charges}.

As is well known in quantum field theory, 
the U(1) gauge symmetries other than $Y$ and $B-L$ are anomalous
with the above contents of chiral fermions.
In the intersecting D-brane models in string theory,
these anomalies are canceled by 
the Green-Schwarz mechanism.
This occurs due to the exchange of the Ramond-Ramond field,
which also makes the anomalous U(1) gauge fields and 
some of the anomaly-free gauge fields massive.
We speculate that a similar mechanism works
in the matrix model as well.

\section{The number of generations}
\label{sec:num-generations}

If all the intersections of 
${\rm S}^2$ and ${\rm S}^2 \times {\rm S}^2$ 
in the 
configuration 
in figure \ref{figure:SM-branes}
%
are similar to (\ref{S2-S2S2-config}),
we get only one generation of the Standard Model fermions.
In section \ref{sec:two-generations}
we show a simple example in which one can get two generations 
by modifying the configuration (\ref{S2-S2S2-config}).
In section \ref{sec:generalization}
we generalize the argument and show that
one cannot get more than two generations
by changing the radii of the ${\rm S}^2$'s in the configuration
or by shifting/rotating the fuzzy manifolds relatively.
In section \ref{sec:three-generations} we show that
three generations can be obtained by squashing one of the 
${\rm S}^2$ of the ${\rm S}^2 \times {\rm S}^2$ in the configuration.


\subsection{Two generations---a simple example}
\label{sec:two-generations}

\begin{figure}[t]
  \begin{center}
  \includegraphics[height=5.5cm]{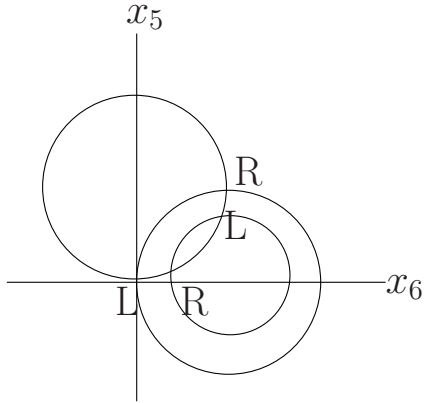}
  \end{center}
  \caption{The intersections of the
${\rm S}^2$ and the ${\rm S}^2 \times {\rm S}^2$,
where the latter is given by 
(\ref{S2S2-eq-2gen}) instead of (\ref{S2S2-eq}).
We have four intersecting points, at which chiral zero modes
appear with the chirality shown by L and R for left-handed 
and right-handed fermions, respectively.
}
  \label{figure:two-gens}
  \end{figure}

In order to get a simple example of configurations giving rise to 
two generations, 
let us multiply a factor $\alpha$
to $L_1$, $L_2$, $\tilde{L}_3$ corresponding 
to the second ${\rm S}^2$ of the fuzzy ${\rm S}^2 \times {\rm S}^2$ 
in (\ref{S2-S2S2-config}).
Then eq.~(\ref{S2S2-eq}) is replaced by
\begin{align}
(x_5 - 1 )^2 + (x_6)^2 + (x_7)^2 = 1  \quad
\mbox{and} \quad 
\left(x_4 - \alpha \right)^2 + (x_8)^2 + (x_9)^2 = 
 \alpha^2 \ .
\label{S2S2-eq-2gen}
\end{align}
For $\alpha < \frac{1}{2}$, 
the ${\rm S}^2$ (\ref{S2-eq})
and the ${\rm S}^2 \times {\rm S}^2$ 
(\ref{S2S2-eq-2gen})
intersect not only 
on $(x_4,x_7,x_8,x_9)= (0 , 0 , 0 , 0 )$
but also on
$(x_4,x_7,x_8,x_9)= (2 \alpha , 0 , 0 , 0 )$.
Slicing the six-dimensional space by these two hyperplanes,
we obtain a view depicted in figure \ref{figure:two-gens}
on the $(x_5,x_6)$-plane.
The small circle that appears from $x_4=2 \alpha$
has a radius $R_{\rm small}=\sqrt{1-4\alpha^2}$, 
and it intersects with another circle when
$1 + R_{\rm small}> \sqrt{2}$, i.e.,\
$\alpha < \left(\frac{\sqrt{2}-1}{2} \right)^{1/2}= 0.455\ldots$
as shown in figure \ref{figure:two-gens}.
In this case, 
the ${\rm S}^2$ and the ${\rm S}^2 \times {\rm S}^2$ 
intersect at four points,
giving rise to two pairs of chiral fermions with opposite chirality.
In what follows we take $\alpha = \frac{1}{8}$ as an example.

\begin{figure}[t]
\begin{center}
\includegraphics[height=5.5cm]{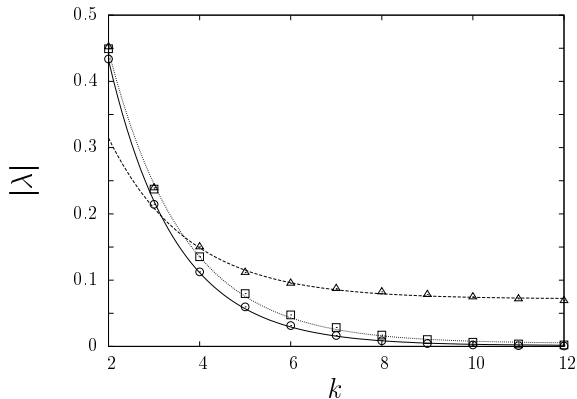}
\end{center}
\caption{
The smallest $|\lambda|$ (circles),
the second smallest one (squares) and
the third smallest one (triangles)
are plotted against $k$
for a configuration analogous 
to (\ref{S2-S2S2-config}) but
corresponding to (\ref{S2S2-eq-2gen}) with $\alpha=\frac{1}{8}$
instead of (\ref{S2S2-eq}).
The solid, dotted and dashed lines 
represent the fits to the behavior 
$|\lambda| = a + b \, {\rm exp}(- c \, k)$.
For the smallest $|\lambda|$, we find 
$a = 0.001(1)$, $b = 1.69(3)$, $c = 0.68(1)$.
%
For the second smallest one, we obtain
$a = 0.004(2)$, $b = 1.49(4)$, $c = 0.61(1)$.
The fitting range for these two cases is $2 \le k \le 12$.
For the third smallest one, we obtain
$a = 0.072(2)$, $b = 0.8(2)$, $c = 0.57(5)$,
where the fitting range is $4 \le k \le 12$.
}
\label{fig:lambda-k-two-gene}
\end{figure}

\begin{figure}[t]
\begin{center}
\includegraphics[height=5.5cm]{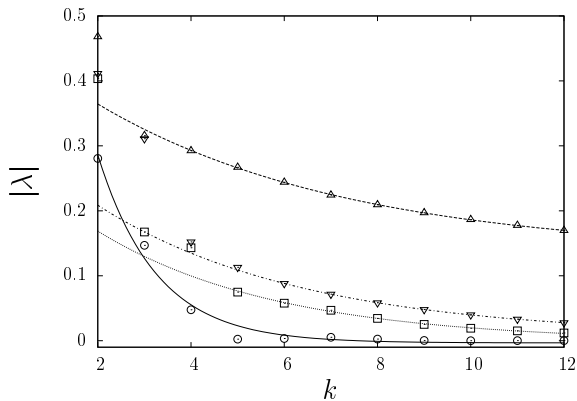}
\end{center}
\caption{
The smallest $|\lambda|$ (circles), the second smallest one (squares),
the third smallest one (inverted triangles)
and the fourth smallest one (triangles)
are plotted against $k$
for a configuration analogous 
to (\ref{S2-S2S2-config}) but
corresponding to (\ref{S2-eq-gen3})
and (\ref{S2S2-eq-gen3})
with $\alpha_2 '=0.35$, $\alpha_2 ''=2$, $\alpha_3=0.38$
instead of (\ref{S2-eq}) and (\ref{S2S2-eq}).
The solid, dotted, dash-dotted and dashed lines 
represent the fits to the behavior 
$|\lambda| = a + b \, {\rm exp}(- c \, k)$.
For the smallest $|\lambda|$, we find 
$a=-0.003(5)$, $b= 1.4(2)$, $c=0.80(7)$ with the fitting range
$2\le k \le 12$.
For the second smallest $|\lambda|$, we find 
$a=-0.001(5)$, $b= 0.29(5)$, $c=0.26(4)$ with the fitting range
$6\le k \le 12$. 
For the third smallest $|\lambda|$, we find 
$a=0.0072(3)$, $b= 0.318(2)$, $c=0.228(2)$ with the fitting range
$6\le k \le 12$. 
For the fourth smallest $|\lambda|$, we find 
$a=0.134(2)$, $b= 0.334(3)$, $c=0.185(5)$ with the fitting range
$4\le k \le 12$.
}
\label{fig:lambda-k-three-gene}
\end{figure}

We solve the eigenvalue equation
(\ref{eigenvalue equation})
for this configuration
and plot the three smallest $|\lambda|$ against $k$
in figure \ref{fig:lambda-k-two-gene}.
We observe that two of them vanish rapidly as $k$ increases,
which suggests the appearance of 
two pairs of chiral zero modes.\footnote{The exponential fit
in figure \ref{fig:lambda-k-two-gene} seems to suggest that 
the second smallest $|\lambda|$ asymptotes to a small but
finite number.
This is also the case with the third 
smallest $|\lambda|$ for the three-generation case
shown in figure \ref{fig:lambda-k-three-gene}.
Whether this leads to a practical problem or not deserves
further investigations.
}

\begin{figure}[t]
\begin{center}
\includegraphics[width=72mm]{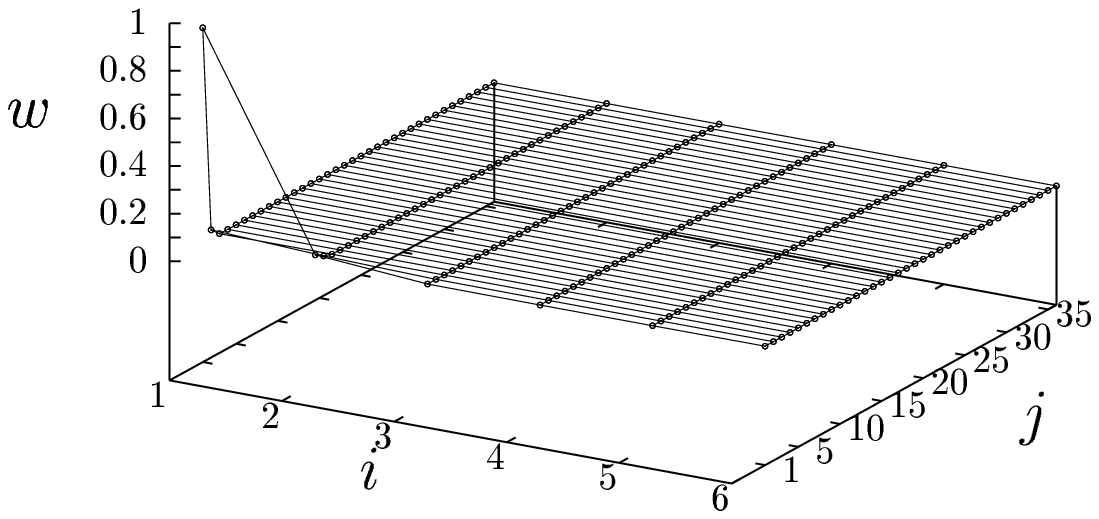}
\includegraphics[width=72mm]{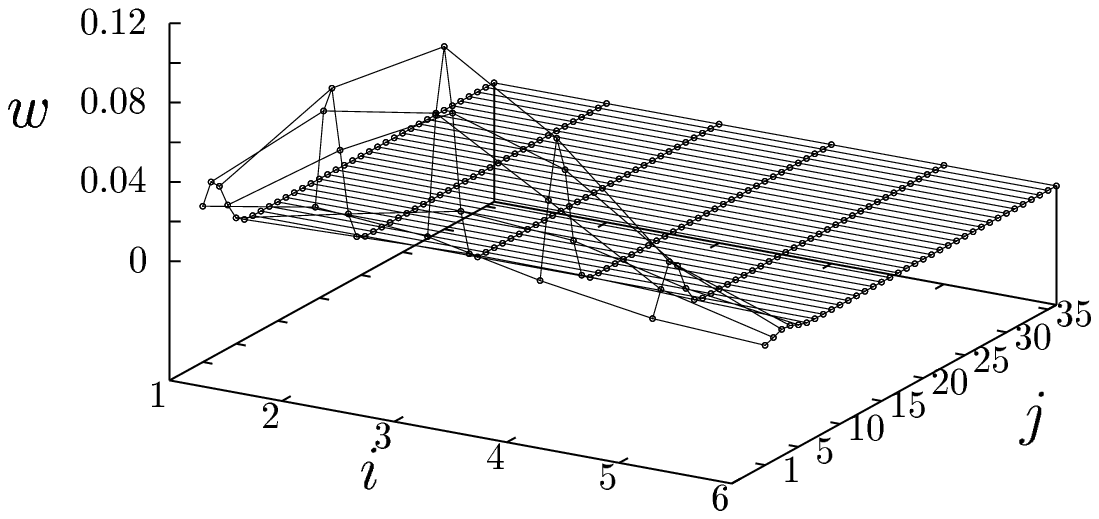}
\includegraphics[width=72mm]{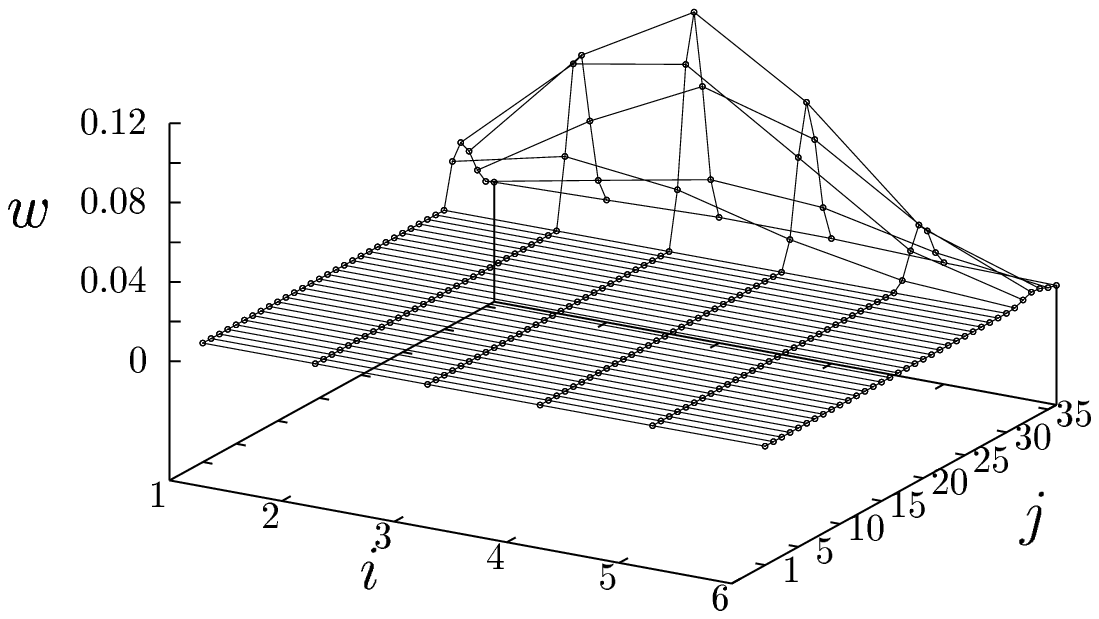}
\includegraphics[width=72mm]{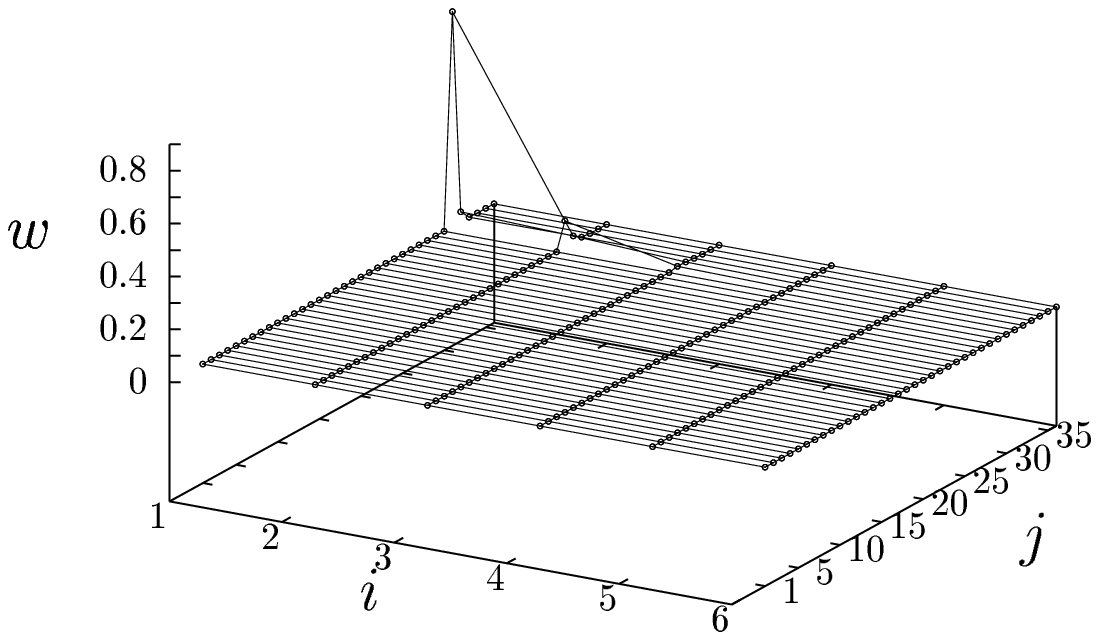}
\end{center}
\caption{
The shape of the wave functions $w(i,j)$ 
of the chiral zero modes with each chirality
found in figure \ref{fig:lambda-k-two-gene} 
is plotted for $k=6$.
On the left (right), we show the results for the left-handed 
(right-handed) modes.
At the top and bottom, we show the results for the
smallest $|\lambda|$ and the second smallest one, respectively.
}
\label{fig:wf-two-gen}
\end{figure}

We take an appropriate linear combination of the two modes
with the smallest $|\lambda|$ so that they have definite chirality
as we have done in eq.~(\ref{phi-LRdecomposed}).
We also do the same thing for 
the two modes with the second smallest $|\lambda|$.
In figure \ref{fig:wf-two-gen}
we show the wave functions of these chiral zero modes.
The plots at the top correspond to the smallest $|\lambda|$,
which look quite similar to the plots in figure \ref{fig:wf}.
These modes are essentially the ones that appeared
in section \ref{sec:fuzzyS2}.
The plots at the bottom correspond to the second 
smallest $|\lambda|$.
These modes are the ones that appear from the 
additional
intersections
using the small circle in figure \ref{figure:two-gens}.
Correspondingly, the wave functions have non-zero values
within the region $31\le j \le 36$, which corresponds to 
the ``north pole'' $(2 \alpha , 0 ,0)$ of 
the second ${\rm S}^2$ of the ${\rm S}^2 \times {\rm S}^2$.
In particular, the right-handed mode is localized
at $(i,j)=(1,31)$ as anticipated.
In the chosen basis, we find that the chiral zero modes that appear
near $(x_5,x_6) =(0,0)$ are seen to be manifestly localized.
As we did in section \ref{sec:fuzzyS2}, we can change the basis
in such a way that the ones that appear near
$(x_5,x_6) = (1,1)$ are seen to be manifestly localized.

By using the warp factor $M$ again,
we can get two generations of the left-handed fermions.
First we restrict ourselves to the form (\ref{M-example}) 
in order to get the left-handed mode 
localized at $(x_5,x_6) =(0,0)$.
Next we change the basis 
in such a way
that the left-handed mode 
appearing near $(x_5,x_6) =(1,1)$
are seen to be localized at $(i,j)=(1,1)$.
Then we require that the warp factor 
\emph{in that basis}
should have the
form (\ref{M-example}).
Note that this requirement is compatible with
(\ref{M-example}) in the original basis
since there are a lot of degrees of freedom
left arbitrary in (\ref{M-example}).
In fact, the number of degrees of freedom in
$M^{(1)}$ and $M^{(2)}$ are
$k^2$ and $(k^2)^2$, respectively,
whereas
the restriction to the form (\ref{M-example})
in a particular basis requires
only $(2k-1)$ and $(2k^2-1)$
for $M^{(1)}$ and $M^{(2)}$, respectively.

\subsection{Generalization}
\label{sec:generalization}

As we have seen above,
the number of generations can be deduced only from 
geometric arguments.
Let us therefore consider
a general configuration of fuzzy ${\rm S}^2$ and
fuzzy ${\rm S}^2 \times {\rm S}^2$, which
can be obtained from the basic configuration (\ref{S2-S2S2-config})
by changing the radii of the ${\rm S}^2$'s that appear
in the configuration
and by shifting/rotating the fuzzy manifolds relatively.
Without loss of generality,
we assume that
the classical manifold corresponding to the fuzzy ${\rm S}^2$ is 
represented by
\begin{align}
(x_6)^2 + (x_4)^2 + (x_5)^2 = 1  \quad
\mbox{and} \quad x_7 = x_8 = x_9 = 0 \ ,
\label{S2-eq-center}
\end{align}
and that corresponding to 
the fuzzy ${\rm S}^2 \times {\rm S}^2$ is represented by
\begin{align}
(\tilde{x}_4)^2 + (\tilde{x}_5)^2 + (\tilde{x}_6)^2 = \rho^2  \quad
\mbox{and} \quad 
(\tilde{x}_7)^2 + (\tilde{x}_8)^2 + (\tilde{x}_9)^2 = \sigma^2 \ .
\label{S2S2-eq-center}
\end{align}
We have introduced 
\begin{align}
\tilde{x}_i = R_{ij}  (x_j - \xi_j) 
\quad \quad \quad 
(i,j = 1 ,\ldots , 6)
\ ,
\label{coord-transf}
\end{align}
where $R \in {\rm SO}(6)$ is a $6\times 6$ matrix representing
a general six-dimensional rotation,
and $\xi$ is a six-dimensional vector,
which represents a general shift.
In order to obtain the intersections, we solve
(\ref{S2-eq-center}) and (\ref{S2S2-eq-center}) simultaneously.
Let us decompose the matrix $R$, the vector $\xi$ and
the six-dimensional coordinate $x$ as
\begin{align}
R  =\left(
\begin{array}{cc}
A & B \\
C & D
\end{array} \right) \ , \quad
\xi =\left(
\begin{array}{c}
a \\
b
\end{array} \right) \ , \quad
x =\left(
\begin{array}{c}
X \\
Y
\end{array} \right) \ .
\label{decompose}
\end{align}
Since $R \in {\rm SO}(6)$, we have a constraint $R^{\rm T} R = {\bf 1}$,
which reads
\begin{align}
A^{\rm T} A + C^{\rm T} C  &= 1  \ ,
\nonumber
\\
B^{\rm T} B + D^{\rm T} D  &= 1 \ , 
\nonumber
\\
A^{\rm T} B + C^{\rm T} D  &= 0 \ .
\label{so6-rel}
\end{align}

In order to obtain the intersections, we may restrict ourselves
to $Y=0$ due to (\ref{S2-eq-center}).
Then we have to solve
\begin{align}
\vec{X}^2 & = 1  \ , 
\label{solve-X-1}
\\
\left(X^{\rm T} - a^{\rm T} , - b^{\rm T} \right)
\left(
\begin{array}{cc}
A^{\rm T} A  & A^{\rm T} B \\
B^{\rm T} A & B^{\rm T} B
\end{array} \right)
\left(
\begin{array}{c}
X - a \\
-b
\end{array} \right)
& = \rho ^2   \ , 
\label{solve-X-2}
\\
\left(X^{\rm T} - a^{\rm T} , - b^{\rm T} \right)
\left(
\begin{array}{cc}
C^{\rm T} C  & C^{\rm T} D \\
D^{\rm T} C & D^{\rm T} D
\end{array} \right)
\left(
\begin{array}{c}
X-a \\
-b
\end{array} \right)
& = \sigma ^2   \ .
\label{solve-X-3}
\end{align}
By adding (\ref{solve-X-2}) and (\ref{solve-X-3})
and using (\ref{so6-rel}), we obtain
\begin{align}
(X - a)^2 + b^2
& = \rho^2 +  \sigma ^2   \ .
\label{solve-X}
\end{align}
From (\ref{solve-X}) and (\ref{solve-X-1}), we obtain
\begin{align}
a \cdot X & = \frac{1}{2} ( a^2 + b^2 + 1 - \rho^2 -  \sigma ^2)   \ .
\label{X-plane}
\end{align}
Thus the problem reduces to solving
(\ref{solve-X-1}), (\ref{solve-X-2}) and (\ref{X-plane}) simultaneously.
The intersection of (\ref{solve-X-1}) and (\ref{X-plane})
gives a circle, while the intersection of 
(\ref{solve-X-2}) and (\ref{X-plane}) gives an ellipse.
Note that a circle and an ellipse on the same plane cannot intersect
at more than four points. 
Therefore, we can obtain only up to two generations 
as far as we restrict ourselves to
a general configuration of fuzzy ${\rm S}^2$ and
fuzzy ${\rm S}^2 \times {\rm S}^2$.

\begin{figure}[t]
  \begin{center}
  \includegraphics[height=5.5cm]{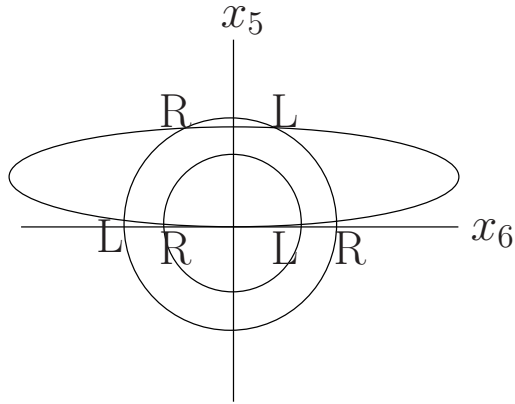}
  \end{center}
  \caption{The intersections of
(\ref{S2-eq-gen3-2}) and (\ref{S2S2-eq-gen3-2}).
We have six intersecting points, at which chiral zero modes
appear
with the chirality shown by L and R for left-handed 
and right-handed fermions, respectively.
}
  \label{figure:three-gens}
  \end{figure}

\subsection{Three generations from squashed ${\rm S}^2$}
\label{sec:three-generations}

In order to get three generations, we need to go beyond
the class of configurations considered in 
section \ref{sec:generalization}.
In general, the dominant background 
can be different from such a configuration
in various ways.
Here we show that
one can actually obtain three generations by squashing one of the 
${\rm S}^2$ of the fuzzy ${\rm S}^2 \times {\rm S}^2$ 
in the configuration.\footnote{It remains to be seen
whether such configurations
with squashed fuzzy spheres
can be realized as solutions to the equation of motion
with possible dynamically generated terms like the Myers term.}
As an example, we consider
a case in which the ${\rm S}^2$ is 
represented by
\begin{align}
(x_4)^2 + (x_5)^2 + (x_6)^2 = 1 \quad
\mbox{and} \quad x_7 = x_8 = x_9 = 0 \ ,
\label{S2-eq-gen3}
\end{align}
and the ${\rm S}^2 \times {\rm S}^2$ is represented by
\begin{align}
\frac{(x_5 - \alpha'_2  )^2}{(\alpha'_2 )^2} 
+ \frac{(x_6)^2}{(\alpha''_2)^2} + (x_7)^2 = 1  \quad
\mbox{and} \quad 
(x_4 - \alpha_3 )^2 + (x_8)^2 + (x_9)^2 = (\alpha_3)^2  \ .
\label{S2S2-eq-gen3}
\end{align}

We solve (\ref{S2-eq-gen3}) and (\ref{S2S2-eq-gen3}) simultaneously.
First we find that $x_4 = 0$ or $2 \alpha_3$.
Then the problem reduces to solving
\begin{align}
 (x_5)^2 + (x_6)^2 & = 1 \quad \mbox{or}
\quad 1  - (2 \alpha_3)^2  \ ,
\label{S2-eq-gen3-2} \\
\frac{(x_5 - \alpha'_2  )^2}{(\alpha'_2 )^2} 
+ \frac{(x_6)^2}{(\alpha''_2)^2}  &= 1  \ ,
\label{S2S2-eq-gen3-2}
\end{align}
simultaneously. 
Eq.~(\ref{S2-eq-gen3-2}) represents two co-centered circles,
while eq.~(\ref{S2S2-eq-gen3-2}) represents
an ellipse.
By choosing the parameters as
\begin{align}
\alpha_2 '  = 0.35 \ , \quad 
\alpha_2 '' = 2 \ ,  \quad 
\alpha_3    = 0.38 \ , 
\label{param-gen3}
\end{align}
we obtain the situation 
depicted in figure \ref{figure:three-gens}
on the $(x_5,x_6)$-plane.

We solve the eigenvalue equation
(\ref{eigenvalue equation})
for this configuration
and plot the four smallest $|\lambda|$ against $k$
in figure \ref{fig:lambda-k-three-gene}.
We observe that three of them vanish rapidly as $k$ increases,
which suggests the appearance of three pairs of chiral zero modes.

\begin{figure}[t]
\begin{center}
\includegraphics[width=72mm]{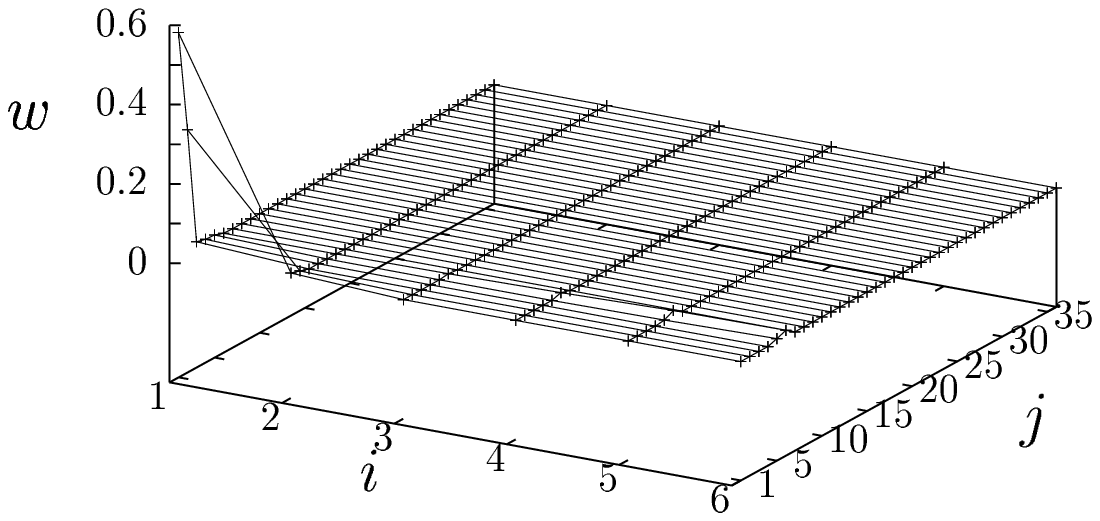}
\includegraphics[width=72mm]{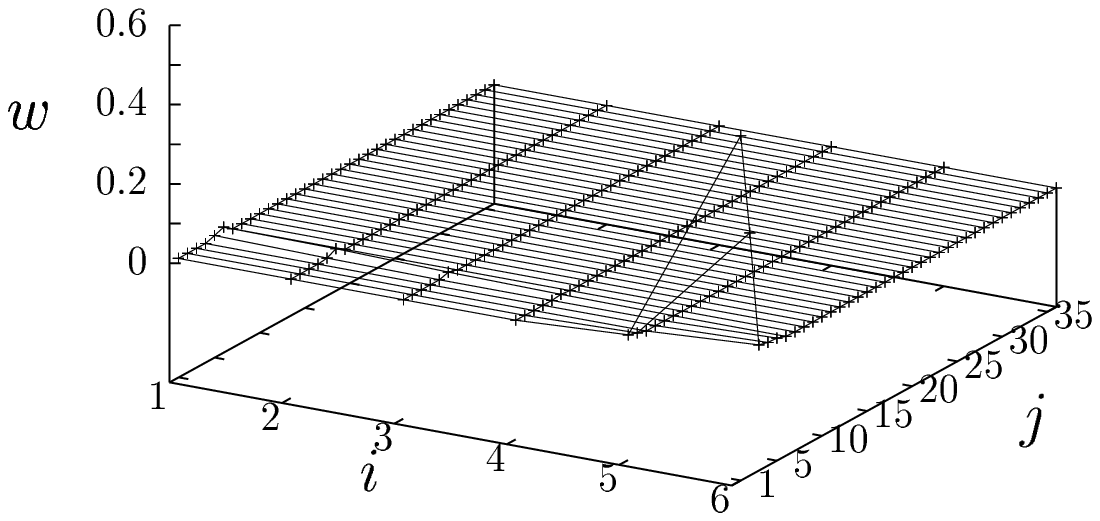}
\includegraphics[width=72mm]{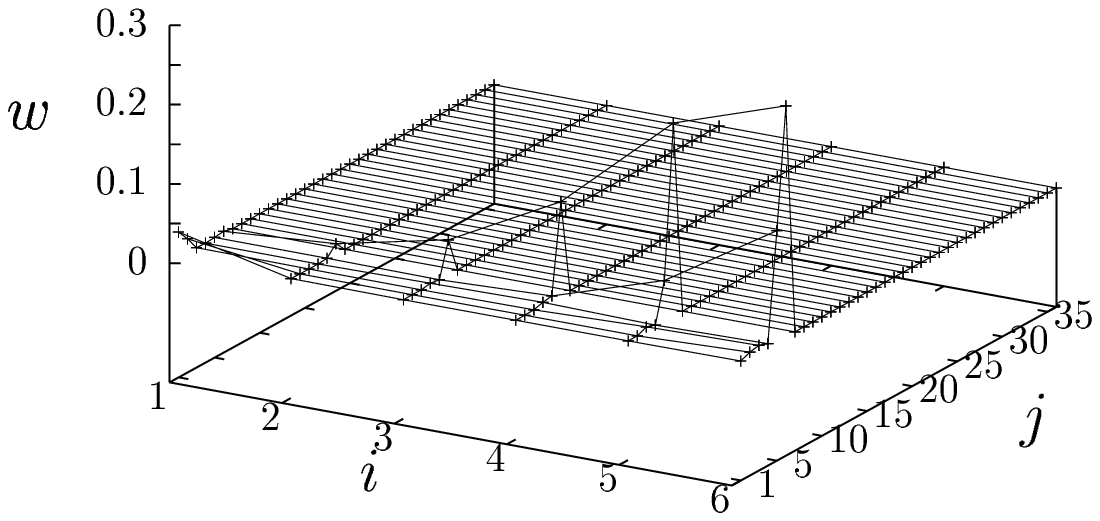}
\includegraphics[width=72mm]{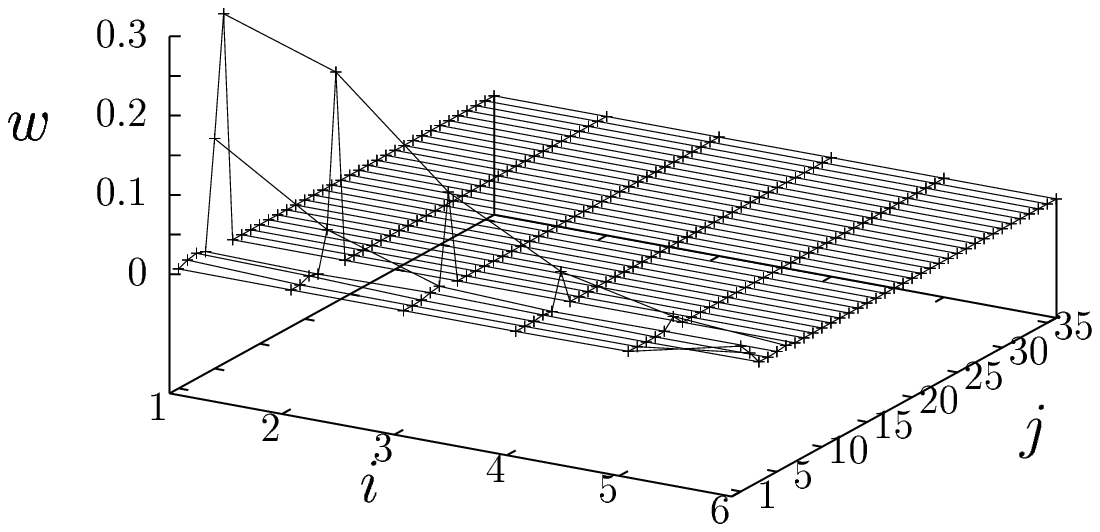}
\includegraphics[width=72mm]{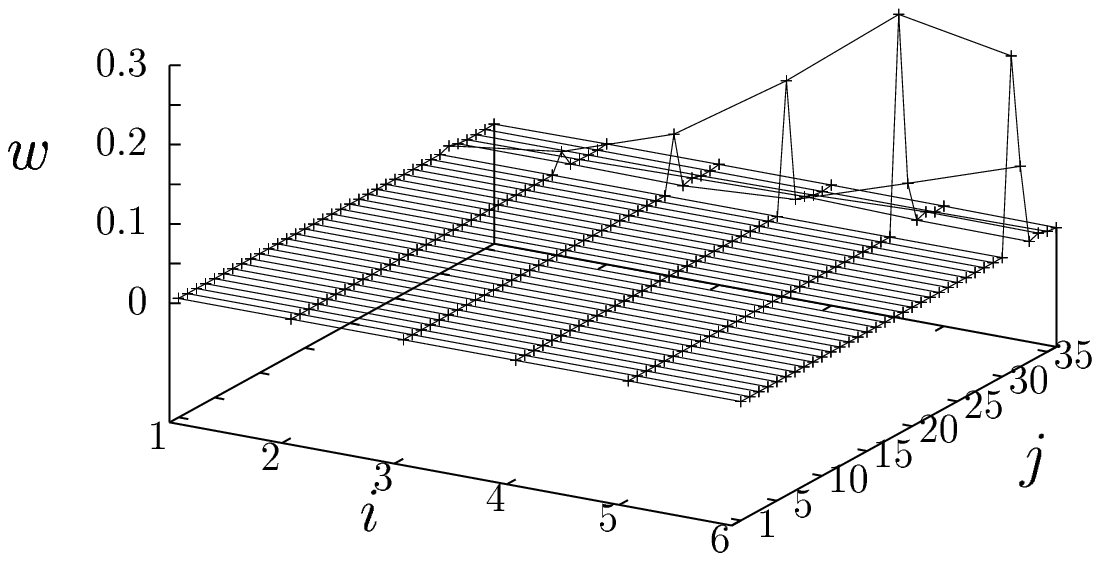}
\includegraphics[width=72mm]{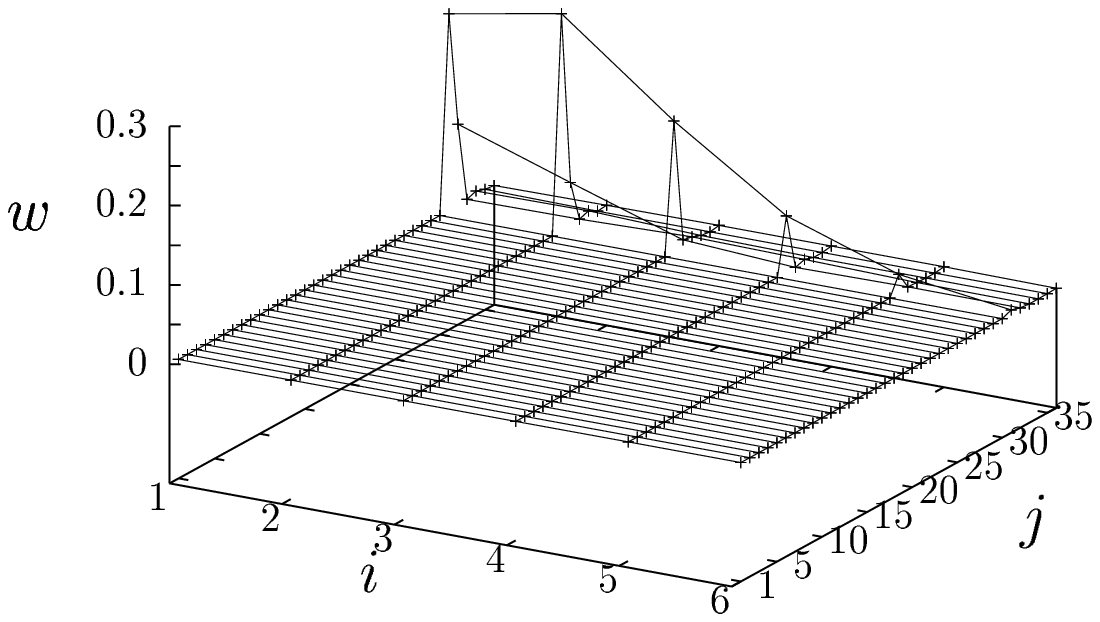}
\end{center}
\caption{
The shape of the wave functions $w(i,j)$ 
of the chiral zero modes with each chirality 
found in figure \ref{fig:lambda-k-three-gene}
is plotted for $k=6$.
On the left (right), we show the results for the left-handed 
(right-handed) modes.
At the top, middle and bottom, we show the results for the
smallest $|\lambda|$, 
the second smallest one and 
the third smallest one, respectively.
}
\label{fig:wf-three-gen}
\end{figure}

We take an appropriate linear combination of the two modes
with the smallest $|\lambda|$ so that they have definite chirality
as we have done in eq.~(\ref{phi-LRdecomposed}).
We also do the same thing for 
the two modes with the second smallest $|\lambda|$,
and similarly for the third smallest $|\lambda|$.
In figure \ref{fig:wf-three-gen}
we show the wave functions of these chiral zero modes.
The plots at the top correspond to the smallest $|\lambda|$.
They are peaked near 
$(i,j)=(1,1)$ and $(i,j)=(6,1)$, respectively.
These are the modes that are localized
at the intersection points
$(x_5,x_6) \sim (0,\pm 1)$ 
in figure \ref{figure:three-gens}.
The plots in the middle row correspond to the second 
smallest $|\lambda|$.
These modes are the ones that appear due to the squashing.
They have non-zero values for $j=6$, which corresponds to the
``north pole'' $x_5\sim 2 \alpha_2 '$ of the first
${\rm S}^2$ of the ${\rm S}^2 \times {\rm S}^2$
and the 
``south pole'' $x_4 \sim 0$ of the second
${\rm S}^2$ of the ${\rm S}^2 \times {\rm S}^2$.
The plots at the bottom correspond to the third smallest $|\lambda|$.
They have non-zero values for $j=31$, which corresponds to the
``south pole'' $x_5\sim 0$ of the first
${\rm S}^2$ of the ${\rm S}^2 \times {\rm S}^2$
and the 
``north pole'' $x_4 \sim 2 \alpha_3$ of the second
${\rm S}^2$ of the ${\rm S}^2 \times {\rm S}^2$.

Similarly to the discussion
at the end of section \ref{sec:two-generations},
we can get three generations of the left-handed fermions
by using the warp factor $M$.
The crucial point is that we can always choose a basis
in such a way that the wave function of each generation
of the left-handed fermion looks like (\ref{phiL-wf}).
The number of arbitrary elements in (\ref{M-example})
is so large that one can impose the condition 
for each generation without conflicts.


\section{Interactions with the gauge field and the Higgs field}
\label{sec:Higgs}

In this section we discuss how the Standard Model fermions
that appear in the model interact with 
the gauge field and the Higgs field.

The gauge field is expected to appear from the fluctuation $a_\mu$
of $A_\mu$ around (\ref{Amu}),
which we decompose as $a_\mu = \tilde{a}_\mu \otimes b$.
Since the gauge field should be a zero mode, we obtain
$[Y_a , [Y_a , b]] = 0 $
in the Lorentz gauge.
This results in a block diagonal $b$ with
$b^{(11)} \propto \1$ and $b^{(22)} \propto \1$
for the explicit example of $Y_a$ in (\ref{S2-S2S2-config}).
The gauge field is therefore insensitive to the wave function 
in the extra dimensions.\footnote{This 
also implies that in the $M=\1$ case the chiral zero
modes with opposite chirality that appear from the same block
interact with the gauge field in the same manner.
Thus we obtain a vector-like gauge theory in that case.
\label{foot:vector-like}
}
This guarantees the universality of the gauge coupling.

Similarly the Higgs field is expected to appear from 
the fluctuation $a_a$ of $A_a$ ($a=4, \ldots , 9$)
around (\ref{Aa}),
which we decompose as $a_a = \tilde{a}_a \otimes b$.
The spectrum of the fluctuation is obtained by
\begin{align}
[Y_a , [Y_a , b]] = \lambda \, b 
\label{Higgs-eq}
\end{align}
in the Lorentz gauge.
As a massless mode, one always has a block diagonal $b$ with
$b^{(11)} \propto \1$ and $b^{(22)} \propto \1$
for the explicit example of $Y_a$ in (\ref{S2-S2S2-config}).
In the case of the 
configuration with five stacks of branes 
in section \ref{sec:SMfermions},
we obtain massless adjoint scalars, which do not couple to the
Standard Model fermions due to the Lorentz symmetry.

On top of this, one can obtain light scalar modes
from the \emph{off-diagonal} block connecting
two of the 
fuzzy ${\rm S}^2$'s
in the configuration when the two spheres come close to each other.
If the ${\rm SU}(2)$ branes come close to the up-type brane,
one obtains a scalar, which has Yukawa couplings to
the SU(2) doublets and the up-type fermions.
Similarly,
if the ${\rm SU}(2)$ branes come close to the down-type brane,
one obtains a scalar, which has Yukawa couplings to
the SU(2) doublets and the down-type fermions.
In the Standard Model, these two scalar fields are
related to each other by G-parity, but in the present case
we obtain them as independent fields 
as in the two-Higgs-doublet model\footnote{See 
ref.~\cite{Branco:2011iw} for a review.
Using their notation,
our situation corresponds to
the type II model
among the four types of the model which avoids 
the tree-level flavor changing neutral current.
This type of 
the two-Higgs-doublet model
has been studied intensively 
since it is the structure that
appears in the Minimal Supersymmetric Standard Model.
}.
If the up-type brane comes close to the down-type brane,
one obtains an SU(2) singlet scalar.
Since the Standard Model fermions localized on these branes
are both right-handed, however, this scalar field does not
couple to the Standard Model fermions due to the Lorentz symmetry.
Note also that since $a_a$ has six components, we have 
six copies of light scalar particles. Their couplings to the
Standard Model fermions are different, however, 
since SO(6) symmetry is broken completely 
by the background configuration.

In general, all the scalar modes acquire mass through radiative
corrections, and decouple from the low-energy spectrum.
We consider that the one that couples most strongly to
top quarks (and right-handed neutrinos)
gets radiative correction to $m^2$ with minus sign,
and eventually induces the electroweak symmetry breaking.
Thus we obtain only the Standard Model Higgs particle
at low energy although we have many more scalar particles
with or without Yukawa couplings to the Standard Model fermions
at high energy.

As an example,
let us consider
the Higgs particle that appears when
the SU(2) branes and the up-type brane come close to each other.
We can calculate the Yukawa couplings
to the quark doublets $Q_{\rm L}$
and the right-handed up-type quarks $U_{\rm R}$
using the three-point coupling
(\ref{fermionic action}) in the type IIB matrix model.
The $3 \times 3$ matrix representing
the Yukawa couplings can be obtained simply from
the overlap of the wave functions as
\begin{align}
\lambda_{a}^{IJ}  & \propto 
S_{a}^{IJ} \equiv
 \tr \left(
(\varphi_{Q_{\rm L}}^{(I)})^\dag  
\Delta_{a} \tilde{b} \, 
\varphi_{U_{\rm R}}^{(J)}  \right) \ ,
\label{Yukawa-calculation}
\end{align}
where $I,J=1,2,3$ represents the generation and
$\Delta_{a}$ represents the $8 \times 8$ gamma matrices in 6d.
The index $a=1,\ldots , 6$ corresponds to the six copies of the Higgs
particles.
The matrix $\tilde{b}$ represents the off-diagonal block 
of the matrix $b$ in (\ref{Higgs-eq})
connecting 
the SU(2) branes and the up-type brane.

If one considers an extreme case in which
the two ${\rm S}^2$ coincide with each other,
(\ref{Higgs-eq}) has a zero mode for 
$\tilde{b} \propto \1$.
%
As far as the two ${\rm S}^2$ are close,
we obtain a light Higgs particle with $\tilde{b} \propto \1$.
Whether we can obtain a realistic structure of 
the Yukawa couplings $\lambda_{a}^{IJ}$ by choosing
the background configuration appropriately
is an interesting open question.

\section{Summary and discussions}
\label{sec:concl}

In this paper we discussed how the Standard Model appears from 
the type IIB matrix model.
While this issue has been discussed by many authors, 
the novelty of our discussion is that we take the model
as it is without any modifications,
and that we consider a constructive
definition starting from finite-$N$ configurations
and taking the large-$N$ limit later on.
In that case,
we have found 
in ref.~\cite{Nishimura:2013moa}
that realizing chiral fermions is much more
difficult than one would expect from the formal arguments
at $N=\infty$.
There it was shown that chiral fermions
in the extra six dimensions should appear in pairs with opposite
chirality. One can, however, introduce a matrix version $M$ of the warp
factor, which enables us to make only the desired chiral fermions
in six dimensions correspond to the ones in four dimensions.
This is always possible by using the huge degrees of freedom in $M$.
It remains to be seen whether the warp factor
determined dynamically has the form required 
for the appearance of chiral fermions.

Accepting this scenario for the appearance
of chiral fermions in the type IIB matrix model,
we discussed whether the Standard Model
appears naturally from the model.
While we basically follow the idea of ref.~\cite{Chatzistavrakidis:2011gs}
using the intersecting branes to obtain chiral zero modes at
the intersections, we have shown that
the Standard Model fermions appear
from quite generic configurations consisting
of fuzzy ${\rm S}^2$ and fuzzy ${\rm S}^2 \times {\rm S}^2$
without fine-tuning.
By virtue of using the two types of fuzzy manifolds
with different dimensionality,
we are able to obtain just the Standard Model fermions plus
the right-handed neutrino.

The Higgs sector is somewhat exotic, though.
We obtain the SU(2) and SU(3) adjoint massless scalars, which 
do not couple to the Standard Model fermions.
We also have possibilities of obtaining two light scalar modes
as SU(2) doublets.
One of them has Yukawa couplings 
to the SU(2) doublets and the up-type fermions,
and the other one has Yukawa couplings 
to the SU(2) doublets and the down-type fermions.
This part resembles the two-Higgs-doublet model \cite{Branco:2011iw}.
Another light scalar can appear as an SU(2) singlet, which
does not couple to the Standard Model fermions.
All these scalar modes appear with multiplicity of six
due to the number of extra dimensions.
We have argued a possibility that the one with 
the strongest Yukawa coupling to the top quark 
(or the right-handed neutrino)
induces the electroweak symmetry breaking
due to radiative corrections to $m^2$,
and survives in the low energy spectrum.

The notion of generations appears naturally from the number
of intersections of a pair of
fuzzy ${\rm S}^2$ and fuzzy ${\rm S}^2 \times {\rm S}^2$.
Unlike in the intersecting D-brane models in compactified space,
however,
the number of generations that can be realized is quite restricted.
We have shown that one obtains
at most two generations for 
a general configuration with 
fuzzy ${\rm S}^2$ and fuzzy ${\rm S}^2 \times {\rm S}^2$.
Three generations can be obtained
by squashing the ${\rm S}^2$ or the ${\rm S}^2 \times {\rm S}^2$,
but other possibilities should certainly be explored.

Since the configuration consisting of 
fuzzy ${\rm S}^2$ and fuzzy ${\rm S}^2 \times {\rm S}^2$
breaks the supersymmetry completely, the hierarchy problem
is an important issue.
Here the hierarchy refers to the one between the electroweak scale
and the Planck scale.\footnote{We would like to mention
that there actually exists 
an even more severe hierarchy problem in Nature,
which is the one between the cosmological constant 
and the Planck scale.
A natural solution to this problem is
suggested within the type IIB matrix model \cite{Kim:2012mw}
based on classical solutions of the model.
}
Among various possibilities proposed in the literature,
the TeV-scale gravity \cite{ArkaniHamed:1998rs}
and the gauge-Higgs 
unification \cite{Manton:1979kb,Fairlie:1979at,Fairlie-2,Hosotani:1983xw,%
Hosotani:1983vn,Hosotani:1988bm,Hatanaka:1998yp}
may be realized in our setup.
In these two scenarios, the existence of extra dimensions
plays a crucial role in explaining the hierarchy.
We consider it interesting that our setup is compatible with
both scenarios.

In the TeV-scale gravity,
one considers that 
the scale of fundamental theory including gravity
is only a few orders of magnitude higher than the TeV scale.
The observed weakness of the gravitational force
is explained by assuming that the extra dimensions
are large, which makes gravity somehow diluted.
Considering that the number of extra dimensions is six
in the present setup, we can have extra dimensions
of the TeV scale if the fundamental theory 
has the scale of $10^3\sim 10^4$ TeV.
Since the fuzzy spheres we discussed
in this paper are expected to appear dynamically
in the matrix model,
we consider it possible that 
they have a radius which is $10^3\sim 10^4$ times larger than 
the fundamental scale of the model.
Thus, our setup fits naturally
into the TeV-scale gravity scenario.

In the gauge-Higgs unification 
scenario,
the Higgs fields are identified as 
extra-dimensional components of the gauge field in 
higher dimensions. 
Then the Higgs mass is protected
from radiative corrections due to the gauge symmetry
in higher dimensions.
As we have discussed in section \ref{sec:Higgs},
the gauge fields and the Higgs fields appear from
$A_\mu$ ($\mu = 0, \ldots , 3$) and $A_a$ ($a=4,\ldots , 9$),
respectively, in the matrix model.
In particular, we are considering a situation in which
the fuzzy ${\rm S}^2$'s, which make up
the SU(2) branes, the up-type brane and the down-type brane,
come close to each other.
When they coincide completely, we obtain four coincident
fuzzy ${\rm S}^2$'s, which give rise to a noncommutative
U(4) gauge theory on ${\rm R}^4 \times {\rm S}^2$ 
generalizing the arguments in ref.~\cite{Iso:2001mg}.
In this way, the Higgs fields are 
identified as extra-dimensional components of 
the gauge field in six dimensions 
although the extra dimensions in our case
have noncommutativity due to the fuzziness of the sphere.
In fact, the fuzzy ${\rm S}^2$'s representing
the SU(2) branes, the up-type brane and the down-type brane
are separated from each other as depicted in 
figure \ref{figure:SM-branes}.
Because of this, the Higgs fields that appear from the off-diagonal
block connecting the two fuzzy ${\rm S}^2$'s
acquire mass, but
the mass is protected 
from radiative corrections due to the gauge symmetry in six dimensions.

There are many open questions.
The most important one is whether the configurations
we considered in this paper can be realized dynamically
in the type IIB matrix model.
As we have discussed in refs.~\cite{Kim:2011ts,Kim:2012mw},
the classical equation of motion is expected to be valid
at late times due to the expansion of the spatial directions.
Therefore, it is expected that the configurations
with a nontrivial structure in the extra dimensions
are realized as a classical solution, possibly 
with quantum corrections (See
refs.~\cite{Imai:2003vr,Imai:2003jb,Kaneko:2005pw}
for related studies in the Euclidean version of the 
type IIB matrix model.)
Another important direction would be 
to calculate the Yukawa couplings from the overlap
of the wave functions 
as we discussed at the end of section \ref{sec:Higgs}.
Of particular interest is to see whether one can reproduce
the experimental data as has been done in closely
related models \cite{Abe:2012fj,Cremades:2004wa}.

Finally, we would like to emphasize that 
the type IIB matrix model has been proposed as
a nonperturbative formulation of superstring theory.
As such, it is expected to be applicable also to cosmology.
From this point of view, we consider that 
the emergence of $(3+1)$-dimensional expanding universe
observed in Monte Carlo simulation \cite{Kim:2011cr}
is remarkable.
More recent work \cite{Ito:2013qga,Ito:2013ywa} suggests
the possibility of reproducing the inflation in the 
early universe from first principle calculations 
in superstring theory.
We hope that the present work provides yet another clue 
to the origin of our universe.

\acknowledgments

We would like to 
thank S.~Iso and Y.~Kitazawa
for valuable discussions.
Computation was carried out on 
supercomputers SR16000 at YITP, Kyoto University 
and FX10 at University of Tokyo.
The authors are supported in part by Grant-in-Aid
for Scientific Research
(No.\ 24540279, 20540286, 24540264, and 23244057)
from JSPS.



\begin{thebibliography}{999}





\bibitem{IKKT}
N.~Ishibashi, H.~Kawai, Y.~Kitazawa, and A.~Tsuchiya,
\emph{A large-N reduced model as superstring},
\emph{Nucl.\ Phys.} {\bf B498} (1997) 467
[{\tt hep-th/9612115}].


\bibitem{Krauth:1998xh}
  W.~Krauth, H.~Nicolai, and M.~Staudacher,
\emph{Monte Carlo approach to M theory},
\emph{Phys.\ Lett.} {\bf B431} (1998) 31
[{\tt hep-th/9803117}].

\bibitem{Austing:2001pk}
  P.~Austing and J.~F.~Wheater,
\emph{Convergent Yang-Mills matrix theories},
\emph{JHEP} {\bf 04 } (2001) 019
[{\tt hep-th/0103159}].

\bibitem{Nishimura:2011xy}
  J.~Nishimura, T.~Okubo and F.~Sugino,
\emph{Systematic study of the SO(10) symmetry breaking vacua 
in the matrix model for type IIB superstrings},
\emph{JHEP} {\bf 1110} (2011) 135
[{\tt arXiv:1108.1293}].

\bibitem{Anagnostopoulos:2013xga}
  K.~N.~Anagnostopoulos, T.~Azuma and J.~Nishimura,
\emph{Monte Carlo studies of the spontaneous rotational 
symmetry breaking in dimensionally reduced super Yang-Mills models},
\emph{JHEP} {\bf 1311} (2013) 009
[{\tt arXiv:1306.6135}].


\bibitem{Kim:2011cr}
  S.~-W.~Kim, J.~Nishimura and A.~Tsuchiya,
\emph{Expanding (3+1)-dimensional universe 
from a Lorentzian matrix model for superstring theory in (9+1)-dimensions},
\emph{Phys.\ Rev.\ Lett.} {\bf 108} (2012) 011601
[{\tt arXiv:1108.1540}].


\bibitem{Aoki:2002jt}
  H.~Aoki, S.~Iso and T.~Suyama,
\emph{Orbifold matrix model},
\emph{Nucl.\ Phys.} {\bf B 634} (2002) 71
[{\tt hep-th/0203277}].


\bibitem{Chatzistavrakidis:2010xi}
  A.~Chatzistavrakidis, H.~Steinacker and G.~Zoupanos,
\emph{Orbifolds, fuzzy spheres and chiral fermions},
\emph{JHEP} {\bf 1005} (2010) 100
[{\tt arXiv:1002.2606}].

\bibitem{Chatzistavrakidis:2012ah}
  A.~Chatzistavrakidis, H.~Steinacker and G.~Zoupanos,
\emph{Orbifold matrix models and fuzzy extra dimensions},
\emph{PoS CORFU} {\bf 2011} (2011) 047
[{\tt arXiv:1204.6498}].


\bibitem{Aoki:2010gv} 
  H.~Aoki,
\emph{Chiral fermions and the standard model 
from the matrix model compactified on a torus}
\emph{Prog.\ Theor.\ Phys.} {\bf 125} (2011) 521 
[{\tt arXiv:1011.1015}].



\bibitem{Abe:2012fj} 
  H.~Abe, T.~Kobayashi, H.~Ohki, A.~Oikawa and K.~Sumita,
\emph{Phenomenological aspects of 10D SYM theory 
with magnetized extra dimensions},
\emph{Nucl.\ Phys.} {\bf B 870} (2013) 30 
[{\tt arXiv:1211.4317}].  


\bibitem{Abe:2013bba}
  H.~Abe, T.~Kobayashi, H.~Ohki, K.~Sumita and Y.~Tatsuta,
\emph{Flavor landscape of 10D SYM theory 
with magnetized extra dimensions},
{\tt arXiv:1307.1831}.


\bibitem{Aoki:2012ei} 
H.~Aoki,
\emph{Probability of the Standard Model appearance from a matrix
model},
\emph{Phys.\ Rev.} {\bf D 87} (2013) 046002 
[{\tt arXiv:1209.4514}].

\bibitem{Aoki:2013pba} 
H.~Aoki,
\emph{Probability distribution over some phenomenological models 
in the matrix model compactified on a torus},
\emph{PTEP} {\bf 2013} (2013) 9, 0903B04
[{\tt arXiv:1303.3982}].


\bibitem{Chatzistavrakidis:2011gs}
  A.~Chatzistavrakidis, H.~Steinacker and G.~Zoupanos,
\emph{Intersecting branes and a standard model realization 
in matrix models},
\emph{JHEP} {\bf 1109} (2011) 115 
[{\tt arXiv:1107.0265}].


\bibitem{Berkooz:1996km}
  M.~Berkooz, M.~R.~Douglas and R.~G.~Leigh,
\emph{Branes intersecting at angles},
\emph{Nucl.\ Phys.} {\bf B480} (1996) 265
[{\tt hep-th/9606139}].

\bibitem{Antoniadis:2000ena}
  I.~Antoniadis, E.~Kiritsis and T.~N.~Tomaras,
\emph{A D-brane alternative to unification},
\emph{Phys.\ Lett.} {\bf B486} (2000) 186
[{\tt hep-ph/0004214}].


\bibitem{Aldazabal:2000sa}
  G.~Aldazabal, L.~E.~Ibanez, F.~Quevedo and A.~M.~Uranga,
\emph{D-branes at singularities: 
a bottom up approach to the string embedding of the standard model},
\emph{JHEP} {\bf 0008} (2000) 002
[{\tt hep-th/0005067}].


\bibitem{hep-th/0007024}
  R.~Blumenhagen, L.~Goerlich, B.~Kors and D.~Lust,
\emph{Noncommutative compactifications of 
type I strings on tori with magnetic background flux},
\emph{JHEP} {\bf 0010} (2000) 006
[{\tt hep-th/0007024}].

\bibitem{Ibanez:2001nd}
  L.~E.~Ibanez, F.~Marchesano and R.~Rabadan,
\emph{Getting just the standard model at intersecting branes},
\emph{JHEP} {\bf 0111} (2001) 002
[{\tt hep-th/0105155}].


\bibitem{Blumenhagen:2001te}
  R.~Blumenhagen, B.~Kors, D.~Lust and T.~Ott,
\emph{The standard model from stable intersecting brane world orbifolds},
\emph{Nucl.\ Phys.} {\bf B616} (2001) 3
[{\tt hep-th/0107138}].

\bibitem{hep-th/0107143}
  M.~Cvetic, G.~Shiu and A.~M.~Uranga,
\emph{Three family supersymmetric 
standard - like models from intersecting brane worlds},
\emph{Phys.\ Rev.\ Lett.} {\bf 87} (2001) 201801
[{\tt hep-th/0107143}].


\bibitem{Cvetic:2001nr}
  M.~Cvetic, G.~Shiu and A.~M.~Uranga,
\emph{Chiral four-dimensional N=1 supersymmetric 
type 2A orientifolds from intersecting D6 branes},
\emph{Nucl.\ Phys.} {\bf B615} (2001) 3
[{\tt hep-th/0107166}].



\bibitem{Cremades:2002dh}
  D.~Cremades, L.~E.~Ibanez and F.~Marchesano,
\emph{Standard model at intersecting D5-branes: Lowering the string scale},
\emph{Nucl.\ Phys.} {\bf B643} (2002) 93
[{\tt hep-th/0205074}].


\bibitem{Kokorelis:2002zz}
  C.~Kokorelis,
\emph{New standard model vacua from intersecting branes},
\emph{JHEP} {\bf 0209} (2002) 029
[{\tt hep-th/0205147}].



\bibitem{Kokorelis:2002qi}
  C.~Kokorelis,
\emph{Exact standard model structures from intersecting D5-branes},
\emph{Nucl.\ Phys.} {\bf B677} (2004) 115
[{\tt hep-th/0207234}].



\bibitem{Steinacker:2013eya}
  H.~Steinacker and J.~Zahn,
\emph{An index for intersecting branes in matrix models},
\emph{SIGMA} {\bf 9} (2013) 067
[{\tt arXiv:1309.0650}].

\bibitem{Nishimura:2013moa}
  J.~Nishimura and A.~Tsuchiya,
\emph{Realizing chiral fermions in the type IIB matrix model at finite N},
\emph{JHEP} {\bf 12} (2013) 002
[{\tt arXiv:1305.5547}].


\bibitem{Myers:1999ps}
  R.~C.~Myers,
\emph{Dielectric branes},
\emph{JHEP} {\bf 9912} (1999) 022
[{\tt hep-th/9910053}].


\bibitem{Imai:2003vr}
  T.~Imai, Y.~Kitazawa, Y.~Takayama and D.~Tomino,
\emph{Quantum corrections on fuzzy sphere},
\emph{Nucl.\ Phys.} {\bf B 665} (2003) 520
[{\tt hep-th/0303120}].


\bibitem{Imai:2003jb}
  T.~Imai, Y.~Kitazawa, Y.~Takayama and D.~Tomino,
\emph{Effective actions of matrix models on homogeneous spaces},
\emph{Nucl.\ Phys.} {\bf B 679} (2004) 143
[{\tt hep-th/0307007}].



\bibitem{Kaneko:2005pw}
  H.~Kaneko, Y.~Kitazawa and D.~Tomino,
\emph{Stability of fuzzy $S^2 \times S^2 \times S^2$
in IIB type matrix models},
\emph{Nucl.\ Phys.} {\bf B 725} (2005) 93
[{\tt hep-th/0506033}].

\bibitem{Steinacker:2014fja}
  H.~C.~Steinacker and J.~Zahn,
\emph{An extended standard model and its Higgs geometry 
from the matrix model},
{\tt arXiv:1401.2020}.



\bibitem{Fukuma:1997en}
  M.~Fukuma, H.~Kawai, Y.~Kitazawa and A.~Tsuchiya,
\emph{String field theory from IIB matrix model},
\emph{Nucl.\ Phys.} {\bf B510} (1998) 158
[{\tt hep-th/9705128}].

\bibitem{Aoki:1998vn}
H.~Aoki, S.~Iso, H.~Kawai, Y.~Kitazawa and T.~Tada,
\emph{Space-time structures from IIB matrix model},
\emph{Prog.\ Theor.\ Phys.} {\bf 99} (1998) 713
[{\tt hep-th/9802085}].



\bibitem{Kim:2011ts}
  S.~-W.~Kim, J.~Nishimura and A.~Tsuchiya,
\emph{Expanding universe as a classical solution 
in the Lorentzian matrix model for nonperturbative superstring theory},
\emph{Phys.\ Rev.} {\bf D 86} (2012) 027901 
[{\tt arXiv:1110.4803}].


\bibitem{Kim:2012mw} 
  S.~-W.~Kim, J.~Nishimura and A.~Tsuchiya,
\emph{Late time behaviors of the expanding universe in the IIB matrix
model},
\emph{JHEP} {\bf 1210} (2012) 147
[{\tt arXiv:1208.0711}].

\bibitem{Nishimura:2012rs} 
J.~Nishimura and A.~Tsuchiya,
\emph{Local field theory from the expanding universe 
at late times in the IIB matrix model},
\emph{PTEP} {\bf 2013} (2013) 043B03 
[{\tt arXiv:1208.4910}].

\bibitem{Branco:2011iw}
  G.~C.~Branco, P.~M.~Ferreira, L.~Lavoura, 
M.~N.~Rebelo, M.~Sher and J.~P.~Silva,
\emph{Theory and phenomenology of two-Higgs-doublet models},
\emph{Phys.\ Rept.} {\bf 516} (2012) 1
[{\tt arXiv:1106.0034}].

\bibitem{ArkaniHamed:1998rs}
  N.~Arkani-Hamed, S.~Dimopoulos and G.~R.~Dvali,
\emph{The hierarchy problem and new dimensions at a millimeter},
\emph{Phys.\ Lett.} {\bf B429} (1998) 263
[{\tt hep-ph/9803315}].




\bibitem{Manton:1979kb}
  N.~S.~Manton,
\emph{A new six-dimensional approach to the Weinberg-Salam model},
\emph{Nucl.\ Phys.} {\bf B158} (1979) 141.



\bibitem{Fairlie:1979at}
  D.~B.~Fairlie,
\emph{Higgs' fields and the determination of the Weinberg angle},
\emph{Phys.\ Lett.} {\bf B82} (1979) 97.


\bibitem{Fairlie-2}
D.~B.~Fairlie,
\emph{Two consistent calculations of the Weinberg angle},
\emph{J.\ Phys.} {\bf G5} (1979) L55.

\bibitem{Hosotani:1983xw}
  Y.~Hosotani,
\emph{Dynamical mass generation by compact extra dimensions},
\emph{Phys.\ Lett.} {\bf B126} (1983) 309.

\bibitem{Hosotani:1983vn}
  Y.~Hosotani,
\emph{Dynamical gauge symmetry breaking as the Casimir effect},
\emph{Phys.\ Lett.} {\bf B129} (1983) 193.


\bibitem{Hosotani:1988bm}
  Y.~Hosotani,
\emph{Dynamics of nonintegrable phases and gauge symmetry breaking},
\emph{Annals Phys.} {\bf 190} (1989) 233.

\bibitem{Hatanaka:1998yp}
H.~Hatanaka, T.~Inami and C.~S.~Lim,
\emph{The gauge hierarchy problem and higher dimensional gauge theories},
\emph{Mod.\ Phys.\ Lett.} {\bf A13} (1998) 2601
[{\tt hep-th/9805067}].

\bibitem{Iso:2001mg}
  S.~Iso, Y.~Kimura, K.~Tanaka and K.~Wakatsuki,
\emph{Noncommutative gauge theory on fuzzy sphere from matrix model},
\emph{Nucl.\ Phys.} {\bf B604} (2001) 121
[{\tt hep-th/0101102}].

\bibitem{Cremades:2004wa}
D.~Cremades, L.~E.~Ibanez and F.~Marchesano,
\emph{Computing Yukawa couplings from magnetized extra dimensions},
\emph{JHEP} {\bf 0405} (2004) 079
[{\tt hep-th/0404229}].


\bibitem{Ito:2013qga}
  Y.~Ito, S.~-W.~Kim, J.~Nishimura and A.~Tsuchiya,
\emph{Monte Carlo studies on the expanding behavior 
of the early universe in the Lorentzian type IIB matrix model},
\emph{PoS LATTICE} {\bf 2013} (2013) 341
[{\tt arXiv:1311.5579}].


\bibitem{Ito:2013ywa}
  Y.~Ito, S.~-W.~Kim, Y.~Koizuka, J.~Nishimura and A.~Tsuchiya,
\emph{A renormalization group method for 
studying the early universe in the Lorentzian IIB matrix model},
{\tt arXiv:1312.5415}.


\end{thebibliography}
\end{document}